\documentclass[11pt, oneside]{article}   	
\usepackage{jheppub}
\usepackage{amsmath}
\usepackage{amssymb}
\usepackage{amsfonts}
\usepackage{amsthm}
\usepackage{amsbsy}
\usepackage{array}
\usepackage{mathtools}
\usepackage{slashed}
\usepackage[usenames,dvipsnames]{xcolor}
\usepackage{subcaption}
\usepackage{dsdshorthand}
\usepackage{graphicx}
\usepackage{ytableau}
\usepackage[vcentermath]{youngtab}

\def\myyoung#1{\,{\scriptsize \young(#1)}\,}
\usepackage{tikz}
\usepackage{bbm}
\usetikzlibrary{snakes}
\usetikzlibrary{decorations}
\usetikzlibrary{trees}
\usetikzlibrary{decorations.pathmorphing}
\usetikzlibrary{decorations.markings}
\usetikzlibrary{external}
\usetikzlibrary{intersections}
\usetikzlibrary{shapes,arrows}
\usetikzlibrary{arrows.meta}
\usetikzlibrary{calc}
\usetikzlibrary{shapes.misc}
\usetikzlibrary{decorations.text}
\usetikzlibrary{backgrounds}
\usetikzlibrary{fadings}
\usetikzlibrary{tikzmark,calc,arrows,shapes,decorations.pathreplacing}
\tikzset{
        cross/.style={cross out, draw=black, minimum size=2*(#1-\pgflinewidth), inner sep=0pt, outer sep=0pt},
	branchCut/.style={postaction={decorate},
		snake=zigzag,
		decoration = {snake=zigzag,segment length = 2mm, amplitude = 2mm}	
    }}

\newcommand{\bea}{\setlength\arraycolsep{2pt} \begin{eqnarray}}
\newcommand{\eea}{\end{eqnarray}}
\def\ft#1#2{{\textstyle{\frac{\scriptstyle #1}{\scriptstyle #2} } }}
\def\fft#1#2{{\frac{#1}{#2}}}

\renewcommand{\(}{\left(}
\renewcommand{\)}{\right)}
\renewcommand{\[}{\left[}
\renewcommand{\]}{\right]}
\newcommand\np{n'}

\def\tinydots{{\resizebox{1em}{!}{$\cdots$}}}
\newcommand{\baa}{\begin{align}}
\newcommand{\eaa}{\end{align}}

\def\lsim{\mathrel{\hbox{\rlap{\lower.55ex \hbox{$\sim$}} \kern-.3em \raise.4ex \hbox{$<$}}}}
\def\gsim{\mathrel{\hbox{\rlap{\lower.55ex \hbox{$\sim$}} \kern-.3em \raise.4ex \hbox{$>$}}}}

\makeatletter
\def\@fpheader{\ }
\makeatother

\title{Gravitational Raman Scattering: a Systematic Toolkit 
for Tidal Effects in General Relativity}
\author{Mikhail M. Ivanov${}^1$, Yue-Zhou Li${}^2$, Julio Parra-Martinez${}^3$, Zihan Zhou${}^2$}
\affiliation{
${}^1$ Center for Theoretical Physics, Massachusetts Institute of Technology, Cambridge, MA 02139, USA \\
${}^2$ Department of Physics, Princeton University, Princeton, NJ 08540, USA  \\
${}^3$ Institut des Hautes Études Scientifiques, 91440 Bures-sur-Yvette, France
}
\emailAdd{ivanov99@mit.edu}
\emailAdd{liyuezhou@princeton.edu}
\emailAdd{julio@ihes.fr}
\emailAdd{zihanz@princeton.edu}
\date{}
\abstract{
We present a framework for systematic 
computations of scattering amplitudes 
for gravitational Raman scattering, -- 
the inelastic scattering of massless
fields off compact relativistic objects. 
We focus on the small-frequency (post-Minkowskian, PM) regime
relevant for the study of tidal effects,
which can be mapped onto 
gravitational wave observables 
during the inspiraling phase of a merger. 
We demonstrate that this setup 
is ideal for systematic 
studies of 
tidal effects, in a way that is free
from coordinate, gauge,
and field redefinition 
ambiguities. 
We use a combination of worldline effective 
field theory, the background field method, and advanced scattering amplitude
techniques to derive phase shifts
for scattering of spin-$0,1,2$ fields off generic compact objects at third PM order. We demonstrate that the inclusion of the recoil of the object is crucial for consistency of this calculation.
Focusing on a particular case 
of black holes, we extract
the leading static and dynamical Love numbers
of the spin-0 field and the static Love number of the spin-1 field in four dimensions 
by matching our EFT amplitudes
and calculations in General Relativity. 
We show, fully on-shell, that the leading static Love numbers vanish 
identically, while the 
dynamical Love numbers 
are not zero and 
run logarithmically.
The latter resolves
the ambiguities of previous off-shell 
matching calculations.
We also extend our results to seven dimensions, 
where spin-2 Love numbers undergo 
a renormalization group running at 2PM, which we compute explicitly. 
In addition, we extract the leading static Love numbers
of spin-0 and spin-1 fields in five dimensions, which also run.
}

\preprint{MIT-CTP/6001}

\begin{document}

\maketitle
\pagenumbering{roman}
\setcounter{page}{2}
\newpage
\pagenumbering{arabic}
\setcounter{page}{1}

\section{Introduction}

The detection of gravitational 
waves by the LIGO/VIRGO/KAGRA
collaboration has started the era of 
precision gravitational wave astronomy~\cite{LIGOScientific:2014pky,VIRGO:2014yos,LIGOScientific:2016aoc,LIGOScientific:2018mvr,LIGOScientific:2020ibl,LIGOScientific:2021usb,LIGOScientific:2021djp,KAGRA:2020agh,LIGOScientific:2025snk}. 
In particular, it has opened up a new channel for probing the internal structure
of gravitational compact  
objects in coalescing binaries, which promises to unveil the detailed structure of black holes and the equation of state of Neutron Stars.
This structure is encoded by
finite-size effects, i.e. the tidal
perturbations of compact bodies, which 
leave measurable imprints 
on the emitted gravitational waves. 
There are two kinds of finite size effects relevant for observations: tidal heating, responsible for the 
absorption of mechanical energy and the
reduction of the emitted flux, 
and conservative tidal deformations, 
which produce distinctive 
contributions 
to the total gravitational 
potential of compact bodies 
that sources the emission 
of the gravitational waves. 

At leading order, the conservative tidal deformations
of compact bodies are captured by
their static tidal Love numbers, which capture body's response to time-independent external 
gravitational fields \cite{PoissonWill2014}. Love numbers 
depend only on the internal 
properties of the gravitating body. 
For neutron stars, they provide important constraints on their equations of state \cite{Flanagan:2007ix,Damour:2009vw}. 
For black holes, 
the static Love numbers vanish identically \cite{Fang:2005qq,Damour:2009vw,Binnington:2009bb,Damour:2009va,Kol:2011vg,Hui:2020xxx,Ivanov:2022hlo,Hadad:2024lsf,Poisson:2020vap,Poisson:2021yau,Riva:2023rcm,Iteanu:2024dvx,Kehagias:2024rtz,Combaluzier-Szteinsznaider:2024sgb,Gounis:2024hcm,Parra-Martinez:2025bcu}, which can be interpreted as a result of hidden symmetries of black hole perturbations~\cite{Charalambous:2021mea,Charalambous:2021kcz,Charalambous:2022rre,Charalambous:2023jgq,Charalambous:2025ekl,Hui:2020xxx,Hui:2021vcv,Hui:2022vbh,Ivanov:2022hlo,Parra-Martinez:2025bcu}. 

The static Love numbers and their time-dependent 
generalizations, dubbed ``dynamical Love numbers,'' are often defined
in the context of the post-Newtonian theory (a formal expansion in body's velocity, or the orbital gravitational potential) (see e.g.~\cite{Binnington:2009bb,Poisson:2014gka,Poisson:2020mdi,Poisson:2020vap,Poisson:2021yau}). This classical definition  is,
however, gauge and coordinate 
dependent. While this raised some 
controversy in the early computations of Love numbers \cite{1986PhRvD..34.3633S,Gralla:2017djj}, these issues had been
later 
resolved at the level of the 
static Love numbers of black holes 
and neutron stars \cite{Damour:2009va,Damour:2009vw,Fang:2005qq,Binnington:2009bb,Poisson:2014gka}. However, generalizing the notion of Love numbers
in general relativity 
beyond the static order proved to be 
a difficult task. The key conceptual 
and computational difficulty is that 
at the high post-Newtonian orders where
the dynamical Love numbers kick in, 
they cannot be unambiguously 
separated from
relativistic corrections
to the point-particle Newtonian potential. 

\begin{figure}
    \centering
    \includegraphics[width=0.5\linewidth]{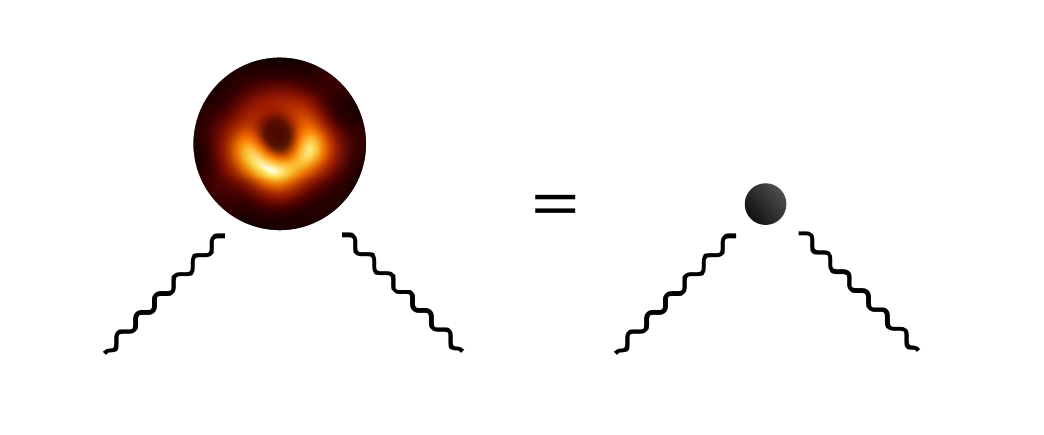}
    \caption{Depiction of the matching of scattering of gravitational waves from a black hole (LHS), and from point particle (RHS).  The black hole picture is adapted from observations by the EHT collaboration~\cite{EventHorizonTelescope:2019dse}.}
    \label{fig:intro}
\end{figure}

A powerful way to overcome the issues with the definition of tidal effects 
and computing their imprints on 
gravitational waveforms is to combine 
point-particle effective field theory
and scattering amplitudes. 
The point-particle effective field theory
(EFT)~\cite{Goldberger:2004jt,Goldberger:2007hy} 
allows one to write a consistent theory 
for perturbative gravity, which respects
the key symmetries, such as gauge 
invariance. In this context 
Love numbers appear as Wilson 
coefficients, $C_i$, of worldline 
curvature couplings. Hence, by construction they are gauge and 
coordinate-independent quantities.
The EFT, however, does not predict their values. The Love numbers need to be matched directly from observational data \cite{Chia:2023tle,Chia:2024bwc,Shterenberg:2024tmo,Andres-Carcasona:2025bni}, or general relativity (GR)
computations, with GR being a UV completion to worldline EFT. Historically, this program started with matching of EFT and GR predictions for static tidal potentials~\cite{Kol:2011vg,Hui:2020xxx,Charalambous:2021mea,Ivanov:2022hlo}. The tidal potential matching is, however, a coordinate and gauge
dependent procedure, which raises concerns about its robustness. 
Scattering events, on the other hand, provide an arena that allows for extracting the Love numbers in a manifestly coordinate, gauge, and field-redefinition invariant manner, by requiring that that amplitudes in GR and EFT are equal
\begin{equation}
    \label{eq:matching}
    {\cal M}^{\rm GR} = {\cal M}^{\rm EFT}(C_i)
\end{equation}
The latter depends on the Wilson coefficients, which are fixed by the
above equality.
This matching procedure is depicted schematically in Fig.~\ref{fig:intro}. 

The combination of EFT and scattering amplitudes methods has previously allowed for an on-shell proof of the vanishing of the static black hole 
Love numbers~\cite{Ivanov:2022qqt},
and also the matching of leading non-zero dissipation coefficients and 
dynamical 
Love numbers~\cite{Saketh:2022wap,Saketh:2023bul}
using the ``near-far'' factorization
of the GR scattering amplitudes. 
More recently, it has been applied in full generality to 
the problem of gravitational Raman scattering, i.e. 
the inelastic scattering of massless field off compact gravitating 
objects  \cite{Ivanov:2024sds,Caron-Huot:2025tlq}. Matching GR and EFT scattering amplitudes for such processes allowed for a consistent matching of Love numbers
of a massless scalar field, including the static and dynamical Love numbers. 
Importantly, the latter exhibits a 
non-trivial renormalization group running,
whose gravitational analog can be probed with gravitational wave data. 
In addition, the imaginary part of 
the Raman scattering amplitude
determined tidal heating effects beyond the leading order, which also undergo 
renormalization through a non-trivial anomalous scaling~\cite{Goldberger:2012kf,Ivanov:2024sds,Ivanov:2025ozg}. This demonstrated that scattering amplitudes enable the extraction of all tidal effects, in a consistent and systematic manner.

In this work, we generalize and extend 
the results of the scalar toy model 
used in~\cite{Ivanov:2024sds} to 
more realistic spin-1 and spin-2 cases, describing the gravitational Raman amplitude of photons and gravitons in EFT. The main goal of our paper 
is to provide a universal toolbox for 
systematic exploration of gravitational 
tidal effects with scattering amplitude 
methods. Our discussion is in general dimension, $D$, which beyond its inherent interest, enables the use of dimensional regularization in the EFT. We also present specific results in $D=4,5,7$.

This paper is structured as follows: In Section~\ref{sec:wlEFT} we review the description of tidal effects in worldline effective field theory, including its proper framing as an open EFT which encodes both conservative and dissipative effects. In Section~\ref{sec:amp} we review various general properties of scattering amplitudes from massless fields from a black hole, including their tensor decomposition, $D$-dimensional partial waves and inversion formulas and exponential representation. In Section~\ref{sec:ampeft} we present the detailed computation of the scattering amplitude in worldline EFT, and the results through 3PM order for scalar, electromagnetic and gravitational waves. In Section~\ref{sec:matching} we extract the leading Love numbers by comparing to full GR results obtained in BHPT. In Section~\ref{sec:comments on RG} we present the 2PM amplitudes in higher dimensions and comment on the running of the static Love numbers in $D=5,7$. Several technical appendices elaborate of some of the points made in the main body of the paper.

\section{
Love numbers and worldline EFTs}
\label{sec:wlEFT}

The response of a compact object to an external force is known to be parameterized by the Love numbers \cite{Damour:2009vw,Damour:1998jk,Goldberger:2004jt,Damour:2009va,Binnington:2009bb,Kol:2011vg}. Tidal forces acting on the compact object induce mass multipole moments. Under the linear approximation, the induced moments are proportional to the tidal moments, with the coefficients defining a linear response function. 

Recall that in the standard materials, such as the Drude model, the response functions in the frequency domain allow smooth Taylor expansions around the zero frequency
\begin{equation} 
G_R(\omega)= \sum_{n=0} c_n (i \omega)^{n}\,,\label{eq: retarded expansion}
\end{equation} 
which include both the conservative response coefficients $c_{2\mathbb{Z}}$ and the dissipative effects captured by $c_{2\mathbb{Z}+1}$. 
The same structures remain in the tidal response. Consider a spherically symmetric and static object, the response functions organized in terms of the eigenbasis of the rotation symmetry, and therefore the response coefficients are labeled by the angular momentum $c_{\ell,n}$. The conservative response coefficients $c_{\ell,2\mathbb{Z}}$ are called the Love number, while $c_{\ell,2\mathbb{Z}+1}$ are the dissipative numbers \cite{Goldberger:2020fot,Ivanov:2022hlo,Saketh:2023bul}.

Nevertheless, the expansion \eqref{eq: retarded expansion} implicitly assumes that the tidal response function can be defined for an isolated system. For a compact object interacting with external perturbations, this assumption breaks down, since the object cannot be cleanly separated from the surrounding radiation. It is therefore necessary to introduce a scale separation and define the tidal response at a given scale. From this perspective, \eqref{eq: retarded expansion} represents a bare response, while the physical response is defined at a finite scale that includes surrounding radiation and picks up contributions from local counterterms.

In addition, there is an ambiguity in computing the tidal response within both Newtonian gravity and general relativity. The tidal response is extracted from the asymptotic expansion of the gravitational potential, which implicitly fixes the gauge of the diffeomorphism (i.e., the choice of coordinates). In this sense, this method of computing the Love numbers suffers from a gauge ambiguity. In the next subsection, we review the worldline EFT approach to the description of the tidal response, which captures the Love numbers covariantly and without gauge ambiguity.

\subsection{The Setup}

To have a gauge-invariant and unambiguous description of both the Love numbers and the dissipative numbers, we use the techniques of the worldline effective field theory (EFT) for compact objects. The spirit of the worldline EFT is to approximate the compact object as a point particle at long distances. This is a coarse-grained picture that arises from ignorance of the internal degrees of freedom of the object. The finite size and dissipative effects leave their imprints by introducing effective multipole operators that non-minimally couple to the worldline. This framework systematically provides gauge-invariant and on-shell perturbative calculations of the tidal response. The Love numbers are unambigiously determined by matching to the GR computations. The worldline EFT include the bulk fields, where we include the massless scalar and photon, as well as gravitational fields
\begin{equation}
S_{\rm bulk}=\int d^D \sqrt{-g}\left(\frac{R}{16\pi G}- \frac{1}{2}(\nabla\phi)^2 -\frac{1}{4} F^2\right)\,.
\end{equation}
The worldline action is schematically given by
\begin{equation}
S_{\rm fs} = -M \int d\tau +
\sum_i 
\int d\tau\, 
Q_\rho^i\cdot\mathcal{T}^\rho_i+S_{\rm fs}^{\rm ct}\,,
\end{equation}
where $\mathcal{T}$ schematically denotes the tidal tensors from either scalar, photon, or gravitational tidal forces, and $i$ running through independent tidal structures such as electric or magnetic responses. $\rho$ denotes the spin indices. $S_{\rm fs}^{\rm ct}$ denotes the counterterm defined on the worldline to absorb the UV divergences, that we will discuss shortly. $Q^{i}_\rho$ denotes the multipole operators that are dressed by the worldline to effectively account for the microscopic details of the compact objects. We have no access to the microscopic details of the compact objects, but the EFT can be defined and enjoy the predictive power by imposing the universal low frequency expansion of the correlators of multipole operators as the retarded response function \eqref{eq: retarded expansion}
\begin{equation}
i \int d\tau e^{-i\omega\tau}\langle[ Q_\rho^{i}(\tau), Q_{\rho'}^{i}(0)]\rangle\theta(\tau)= \delta_{\rho\rho'}G_{R,\ell}^{i}(\omega)=\delta_{\rho\rho'}\big(c_{\ell,0}^{i}+ic_{\ell,1}^{i}\omega+c_{\ell,2}^{i}\omega^2+\cdots\big)\,.\label{eq: tidal response}
\end{equation}
It is worth noting that this is the bare response function, approximately treated as if the system were isolated, such as a collection of oscillators. A more refined picture is that the multipoles are not isolated but rather dragged by external gravitational perturbations. Therefore, the response functions need to be renormalized to be physical. We will discuss this point in more detail later. 

Note also that both the multipoles and the tidal tensors describe transverse fluctuations with respect to the worldline. 
This suggests that they are labeled by ${\rm SO}(D-1)$ that preserves the worldline motion. Thus, $\rho$ labels the irreducible representation of ${\rm SO}(D-1)$, and the indices $(a_1 \cdots a_L)$ run over the transverse directions. The tidal tensors are then constructed by acting spatial derivatives on the modes of dynamical fields that either align or perpendicular to the worldline motion. To be more precise, we have

\begin{itemize}
\item Scalar
\begin{equation}
T_{\phi}^{a_1\cdots a_L}=P_{\mu_1\cdots\mu_L}^{a_1\cdots a_L}\nabla^{(\mu_1}\cdots \nabla^{\mu_L)}\phi ~,
\label{eq: scalar tides}
\end{equation}
\item Photon
\begin{align}
& T_{\gamma,B}^{a_1\cdots a_L}= P_{\mu_1\cdots\mu_L}^{a_1\cdots a_L}\nabla^{(\mu_1}\cdots \nabla^{\mu_{L-2}}F^{\mu_{L-1})\mu_{L}}\,,\nonumber\\
& T_{\gamma,E}^{a_1\cdots a_L}= P_{\mu_1\cdots\mu_L}^{a_1\cdots a_L}\nabla^{(\mu_1}\cdots \nabla^{\mu_{L-1}}F_\nu\,^{\mu_L)} u^\nu\,.\label{eq: photon tides}
\end{align}
\item Graviton
\begin{align}
& T_{h,T}^{a_1\cdots a_L}=P_{\mu_1\cdots\mu_L}^{a_1\cdots a_L}\nabla^{(\mu_1}\cdots \nabla^{\mu_{L-4}} C^{\mu_{L-3}|\mu_{L-2}|\mu_{L-1})\mu_L}\,,\nonumber\\
& T_{h,B}^{a_1\cdots a_L}=P_{\mu_1\cdots\mu_L}^{a_1\cdots a_L}\nabla^{(\mu_1}\cdots \nabla^{\mu_{L-3}} C^{\mu_{L-2}|\mu_{L-1}|\mu_{L})}\,_\nu u^\nu\,,\nonumber\\
& T_{h,E}^{a_1\cdots a_L}=P_{\mu_1\cdots\mu_L}^{a_1\cdots a_L}\nabla^{(\mu_1}\cdots \nabla^{\mu_{L-2}} C^{\mu_{L-1}}\,_\nu\,^{\mu_{L})}\,_\rho u^\nu u^\rho\,.\label{eq: graviton tides}
\end{align}
\end{itemize}

The projector $P_{\mu_1 \cdots \mu_L}^{a_1 \cdots a_L} = \prod_i e_{\mu_i}\,^{a_i}$ is constructed using the local tetrad and projects the indices onto the spatial plane orthogonal to the worldline motion, where $(\mu_1 \cdots \mu_L)$ denotes traceless symmetrization, and $|\mu|$ denotes the index that is not included in the symmetrization. It is worth emphasizing that $L$ counts the number of free transverse indices (equivalently, the total number of boxes in the associated Young diagram). The ${\rm SO}(D-1)$ representation $\rho$ is determined by the shape of the Young diagram, not by $L$ alone. We explicitly show the tensor structure as follows:
\ytableausetup{
  boxsize=2em,
  centertableaux,
  mathmode
}
\begin{equation}
    \begin{aligned}
        & T_{\phi}^{a_1\cdots a_L}: \quad         \small{\begin{ytableau}
            a_1 & a_2 & \ldots & a_{L}
        \end{ytableau}}\\
        & T_{\gamma,B}^{a_1\cdots a_L}: \quad 
         \small{\begin{ytableau}
            a_1 & a_2 & \ldots & a_{\smash{L\!-\!1}}\\
            a_L
        \end{ytableau}}
        ~, 
        \quad
        T_{\gamma,E}^{a_1 \cdots a_L}: \quad 
        \small{\begin{ytableau}
            a_1 & a_2 & \ldots & a_L
        \end{ytableau}} \\
        & T_{h, T}^{a_1\cdots a_L}: \quad
        \small{\begin{ytableau}
            a_1 & a_2 & \ldots & a_{\smash{L\!-\!2}} \\
            a_{\smash{L\!-\!1}} & a_{L}
        \end{ytableau}} ~, \quad
        T_{h,B}^{a_1 \cdots a_L}: \quad 
        \small{\begin{ytableau}
            a_1 & a_2 & \ldots &  a_{\smash{L\!-\!1}} \\
            a_L
        \end{ytableau}} ~, \quad
        T_{h,E}^{a_1\cdots a_L}: \quad
         \small{\begin{ytableau}
            a_1 & a_2 & \ldots & a_L
        \end{ytableau}} ~.
    \end{aligned}
\end{equation}

Let us comment on how our notations are consistent with more standard conventions, especially in $D=4$, such as \cite{Hui:2020xxx,Charalambous:2021mea}. As in standard electromagnetic dynamics, $F_{0}\,^a = F_{\mu}\,^a u^\mu = E^a$ represents the electric field. Therefore, $T_{\gamma,2}^{a}$ is referred to as the electric part, constructed from the electric fields. On the other hand, $F^{ab} := B^{ab}$ denotes the magnetic field, defining $T_{\gamma,1}^{ab}$ as the magnetic tensors. In $D=4$, the magnetic field is conventionally defined as $B^a = \epsilon^{abc} B_{bc}$. To align with this 4D convention, we simply need to redefine the magnetic multipoles by, e.g., $Q^{ab}=\epsilon^{abc}Q_c$, such that we have $Q^a B_a$ instead in $S_{\rm fs}$. Similarly for graviton, the standard notation is to call $C^{a0b0}=E^{ab}$ the electric Weyl. $C^{0abc}=B^{abc}$ is the magnetic part and it translates to 4D convention by $H^{ab}=-1/2 \epsilon^{0acd}B_{cdb}$ and redefining $Q$ appropriately. It is notable that $T_{h,1}$ is a new tensor appearing in $D>4$, which identically vanishes in $D=4$.

\subsection{Schwinger-Keldysh open EFT}

Physically, we aim to describe the response to tidal forces. One can therefore integrate out the internal operators using linear response theory, which generates higher-derivative corrections to the worldline action. However, because dissipation is present, the system is intrinsically open. To capture both tidal response and dissipation, it is convenient and instructive to formulate the EFT using the Schwinger–Keldysh (SK) formalism \cite{Schwinger:1960qe,keldysh2024selected}.

The SK formalism requires doubling the fields on two timefolds
\be
S_{\rm fs}^{\rm SK}=S_{\rm fs}^1-S_{\rm fs}^2\,.
\ee
We shall use the Keldysh basis
\be
\mathcal{T}^c=\fft{\mathcal{T}^1+\mathcal{T}^2}{2}\,,\quad \mathcal{T}^r=\mathcal{T}^1-\mathcal{T}^2\,,
\ee
which explicitly distinguish the classical field $\mathcal{T}^c$ and its response field $\mathcal{T}^r$. Therefore, the relevant interaction terms are
\be
\int d\tau \left(Q^r \mathcal{T}^c+Q^c \mathcal{T}^r\right)\,.
\ee
We now systematically integrate out the multipole operators
\be
\int DQ^c DQ^r e^{i\int d\tau \left(Q^r\mathcal{T}^c+Q^c\mathcal{T}^r\right)} \sim 
{\color{red} \underbrace{ \color{black} e^{i \int d\tau d\tau' G_R(\tau,\tau')\mathcal{T}^r(\tau)\mathcal{T}^c(\tau')} }_{Z_R}} 
\times 
{\color{red} \underbrace{ \color{black} e^{-\frac{1}{2} \int d\tau d\tau' G_K(\tau,\tau')\mathcal{T}^r(\tau)\mathcal{T}^r(\tau')} }_{Z_K}}
\ee
where $G_K(\tau,\tau')=\langle Q^c(\tau) Q^c(\tau')\rangle= \langle \{Q(\tau),Q(\tau')\}\rangle$ is the Keldysh (symmetric) correlator \cite{Liu:2018kfw}. In the EFT regime, we insert \eqref{eq: tidal response} for the retarded Green function $G_R$, and all Love numbers and dissipation coefficients in Eq.~\eqref{eq: tidal response} are treated as Wilson coefficients encoded in $Z_R$. The Keldysh correlator can then be determined by applying the fluctuation–dissipation theorem to $G_R$, which in the Boulware vacuum reads\footnote{For other vacua, such as the Unruh vacuum and Hartle-Hawking vacuum, see \cite{Goldberger:2019sya,Goldberger:2020geb}.}
\be
G_K(\omega)=2 {\rm Im}\, G_R(\omega)\,.
\ee
Since our motivation is to understand the tidal response, in this paper we only keep $Z_R$. Nevertheless, we emphasize that $Z_K$ is important for extracting the universal RG behavior of the multipole operators that capture the universal tails of the gravitational waveform \cite{Ivanov:2025ozg}.

We substitute the low-frequency expansion \eqref{eq: tidal response} into $Z_R$ and then study the consequences of the resulting effective action. Notably, we can further distinguish a conservative sector and a dissipative sector, $Z_R = Z_R^{\rm conser} Z_R^{\rm diss}$, where the former encodes the tidal Love numbers and the latter encodes the dissipative coefficients. We will now discuss them separately.

\subsubsection{Conservative sector}

We now focus on the real part of the response function (for real frequencies) in \eqref{eq: tidal response}, which captures the conservative finite-size effects and encodes the Love numbers. We focus on
\begin{equation}
Z_R^{\rm conser}= e^{i \int d\tau d\tau' {\rm Re}\,G_R(\tau,\tau') \mathcal{T}^r(\tau)\mathcal{T}^c(\tau')}\,.
\end{equation} 
This generates higher-dimensional operators that capture the finite-size effects. These higher-dimensional operators are precisely those that construct all finite-size counterterms. Therefore, the conservative parts of $G_{R,\ell}^{s,i}$ can all be absorbed into these counterterms. The low-lying terms read
\begin{align}
& S_{\rm fs}^{\rm ct}\big|_{\phi}=\fft{1}{2} \int d\tau \big( C_{1}^\phi (\boldsymbol{\partial}_\mu \phi)^2+C_{\omega^2 0}^\phi \dot{\phi}^2  + \cdots\big)\,,\nn\\
& S_{\rm fs}^{\rm ct}\big|_{\gamma}=\fft{1}{2} \int d\tau \big(C^{\gamma}_{E,0} E_\gamma^2+\fft{1}{2}C^{\gamma}_{B,0} B_\gamma^2+\cdots\big)\,,\nn\\
& S_{\rm fs}^{\rm ct}\big|_h=\fft{1}{4}\int d\tau\big(C^{h}_{E,0} E_h^2+\fft{1}{2} C^{h}_{B,0} B_h^2+ \fft{1}{4} C^{h}_{T,0}T_{h,1}^2\cdots \big)\,,\label{eq: fs-ct-few}
\end{align}
where the dots refer to higher-derivative terms with either spatial or time derivatives, and we define $\boldsymbol{\partial}=\left(g^{\mu\nu}+u^\mu u^\nu\right)\partial_\mu$ and $\dot{\phi}=u^\mu\partial_\nu \phi$. We thus have $\log Z_R^{\rm conser}=i\left(S_{\rm fs}^{{\rm ct},+}-S_{\rm fs}^{{\rm fs},-}\right)$.

It is worth noting that the Wilson coefficients in $S_{\rm fs}^{\rm ct}$ are re-label of those in \eqref{eq: tidal response}. Translating between these conventions is straightforward, as we only need to compare \eqref{eq: fs-ct-few} with the action obtained from “squaring” the tidal tensors in \eqref{eq: scalar tides}, \eqref{eq: photon tides}, and \eqref{eq: graviton tides}.

As we will discuss in section \ref{subsec: retarded vs Feynman}, we primarily work with positive frequencies, where the retarded propagator coincides with the Feynman propagator in the Boulware vacuum. Therefore, for the conservative sector, a tremendous simplification arises because we are allowed to work directly with the action \eqref{eq: fs-ct-few}, as in the usual in–out EFT, to compute amplitudes. This allows us to apply amplitude-organizing principles and modern techniques. For example, we can show that \eqref{eq: fs-ct-few} precisely provides gauge-invariant polarization structures in worldline amplitudes. This is directly analogous to the case of standard four-point amplitudes in QFTs, as demonstrated in \cite{Ivanov:2024sds}, where the polarization structures are effectively captured by the leading higher-derivative terms in the Lagrangian. The phase shifts or amplitudes of the worldline EFT can then be factorized into these polarization structures, multiplied by analytic functions of the frequency and scattering angle, up to physical branch cuts and simple poles. In $D=4$, these polarization structures can be neatly described using the helicity formalism \cite{Elvang:2013cua}.

\subsubsection{Dissipative sector}

Terms with the odd powers of frequency 
in the response function 
describe non-conservative 
effects such as the dissipation 
This phenomenon can be straightforwardly discussed in SK formalism in $Z_R$. In particular, these effects are encoded by the dissipative Wilson coefficients in $Z_R^{\rm diss}$. For example, for the leading dissipation we have
\be
& S_{R}^{\rm diss}\big|_{\phi}=\int d\tau\left(C_{\omega0}^\phi \phi^c \dot{\phi}^q+C_{\omega1}^\phi \boldsymbol{\partial}\phi^c\cdot \boldsymbol{\partial} \dot{\phi}^q +\cdots\right)\,,\nn\\
& S_{R}^{\rm diss}\big|_{\gamma}= \int d\tau\left(C_{E,\omega 0}^\gamma E_\gamma^c \cdot \dot{E}_\gamma^q+\fft{1}{2}C_{B,\omega 0}^\gamma B_\gamma^c \cdot \dot{B}_\gamma^q+\cdots\right) \,,\nn\\
& S_{R}^{\rm diss}\big|_{h}= \int d\tau\left(C_{E,\omega0}^h E_h^c \cdot \dot{E}_h^q+\fft{1}{2}C_{B,\omega0}^h B_h^c \cdot \dot{B}_h^q+\fft{1}{4} C_{T,\omega0}^h T_{h,1}^c \cdot T_{h,1}^q\cdots\right) 
\ee

These dissipation effects are essentially the absorption of fields by a compact object. To see this, we can consider the simplest absorption process: a transition 
from an initial state $|i\rangle$ consisting of the ground state
of the worldline 
plus a graviton, to an excited
final state of the worldline $\langle f|$ without
a graviton. The inclusive absorption cross section of this process is computed by the Wightman function from squaring the amplitudes and summing over $f$
\be
\sigma_{\rm abs}=\fft{1}{2\omega} \int d\tau \langle Q(\tau)Q(0)\rangle e^{-i\omega\tau}\,.
\ee
In the Boulware vacuum, the fluctuation-dissipation theorem dictates 
\be
\sigma_{\rm abs}=\fft{G_{W}^{>}(\omega)}{2\omega}=\fft{{\rm Im}\, G_R(\omega)}{\omega}\,.
\ee
As we mention, in the Boulware vaccum, $G_R$ is essentially the in-out Feynman Green function for positive frequency. Therefore, the advantage of this expression is that we can  calculate radiative corrections in the convenient in-out approach and then translate the results into the in-in correlators relevant for the absorption cross-sections. In particular, thanks to the universality of EFT, the wavefunction renormalization condition extracted from the in-out observables also works for these in-in observables.

\section{Properties of the scattering amplitude}
\label{sec:amp}

In this section we describe various general properties of the scattering amplitude, which we will later use in our computation of gravitational Raman scattering in EFT.

\subsection{Tensor structures}

For photon and graviton scattering off the worldline, the amplitudes can be conveniently factorized as
\begin{equation}
\mathcal{M} = \sum_{(i)}{\rm Poly}^{(i)}\big((p_j,\epsilon_j)\big) A^{(i)}(\omega,\cos\theta)\,,
\end{equation}
where ${\rm Poly}^{(i)}\big((p_j,\epsilon_j)\big)$ are polynomials of momenta and polarizations, which are transverse and traceless, and encode the gauge redundancies
\begin{equation}
p_j^2=\epsilon_j^2=p_j\cdot\epsilon_j=0\,,\quad \epsilon_j^\mu \simeq \epsilon_j^\mu+ \mathbb{C}\, p_j^\mu\,,
\end{equation}
and the coefficients $A^{(i)}$ are pure functions of frequency and the scattering angle, which effectively behave as spin-0 amplitudes.

The tree-level amplitudes of photon and graviton computed from Eq.~\eqref{eq: fs-ct-few} can serve as a natural basis for the polarization structures, ${\rm Poly}^{(i)}$. For the simplest scalar case, the leading tidal effects come from Eq.~\eqref{eq: fs-ct-few}
\begin{equation}
\label{eq:phi_tidal}
    \mathcal{M}_\phi = \omega^2 (C_1^\phi +C^\phi_{\omega^20}) - 2 C^\phi_{1} \omega^2 x^2 ~.
\end{equation}
For photon case, we consider
\begin{equation}
\mathcal{M}_\gamma = H_{12} A^{(1)}_\gamma + V_{12} A^{(2)}_\gamma\,,\label{eq: basis A}
\end{equation}
where
\begin{equation}
H_{12}=\frac{1}{\omega^2}F^{1}_{\mu\nu}F^{2\mu\nu}\,,\quad V_{12}=-\frac{4}{\omega^2} (F^1)_\mu\,^\nu (F^2)_{\nu\rho} u^\mu u^\rho +H_{12}\,.\label{eq: bas photon}
\end{equation}
In these expressions, $(F^i)_{\mu\nu}$ is kinematic structure that can be viewed as built from the on-shell U$(1)$ field
\begin{equation}
(F^i)_{\mu\nu}=i (p_{i\mu}\epsilon_{i\nu}-p_{i\nu}\epsilon_{i\mu})\,.
\end{equation}
Thus Eq.~\eqref{eq: bas photon} is completely captured by the action Eq.~\eqref{eq: fs-ct-few}, for example 
\begin{equation}
\label{eq:photon_tidal}
\mathcal{M}_\gamma^{\rm ct}=\fft{1}{4}\omega^2\big(C_{B,0}^{\gamma}+C_{E,0}^{\gamma}\big)H_{12}+\fft{1}{4}\omega^2\big(C_{B,0}^{\gamma}-C_{E,0}^{\gamma}\big)V_{12}+\cdots\,.
\end{equation}

The polarization basis of graviton case can be constructed by squaring the photon polarizations, namely
\begin{equation}
\mathcal{M}_h = H_{12}^2 A^{(1)}_h +V_{12}^2 A^{(2)}_h-\frac{1}{4} H_{12}V_{12} A^{(3)}_h\,.\label{eq: basis h}
\end{equation}
This is a toy example of the double copy relation \cite{Bern:2008qj,Kawai:1985xq}.
The last structure is simply the Gram determinant of kinematic variables
\begin{align}
\fft{{\rm det}\,{\rm Gram}(u,p_1,p_2,\epsilon_1,\epsilon_2)}{\omega^4}=-\frac{1}{4} H_{12}V_{12}:=\mathcal{G}\,,\label{gram det}
\end{align}
which vanishes identically in $D=4$.

Again, we realize that these structures are captured by the action in \eqref{eq: fs-ct-few}
\begin{align}
& C^1_{\mu\nu\rho\sigma}C^{2\mu\nu\rho\sigma}=\frac{1}{4}H_{12}^2 \omega^4\,,\quad C^1_{\mu\nu\rho\sigma}C^{2\mu\nu\rho}\,_\alpha u^\sigma u^\alpha=-\frac{1}{16}H_{12}\big(H_{12}-V_{12}) \omega^4\,,\nonumber\\
& C^1_{\mu\nu\rho\sigma}C^{2\mu}\,_\alpha\,^\rho\,_\beta u^\nu u^\sigma u^\alpha u^\beta=\frac{1}{64}(H_{12}-V_{12})^2 \omega^4\,.
\end{align}
We then have, for example
\begin{equation}
\begin{aligned}
\label{eq:grav_ct}
\mathcal{M}_h^{\rm ct}& =\frac{1}{512} \omega^4 (3C_{T,0}^h-3C_{B,0}^h+4C_{E,0}^h)H_{12}^2 +\frac{1}{512}\omega^4(3C_{T,0}^h+3C_{B,0}^h+4C_{E,0}^h)V_{12}^2 \\
& \quad +\frac{1}{256}\omega^4(3C_{T,0}^h-4 C_{E,0}^h)H_{12}V_{12}+\cdots\,.
\end{aligned}
\end{equation}

\subsection{Partial wave expansion} 
\label{subsec:partial_wave}

We will now explain our construction of the partial waves for gravitational Raman scattering, which closely follows Ref.~\cite{Caron-Huot:2022jli}, and the extraction of partial wave coefficients. The discussion is purely based on rotational symmetry, so it is the same in full GR and worldline EFT, but we will use the worldline language for convenience. Our construction is fully covariant by using the velocity of the black hole or worldine $u^\mu$ as a reference vector, which allows for a convenient decomposition in terms of the polarization structures described in the previous subsection. To lighten the technical load of this section, some details are relegated to Appendix \ref{app: partial wave}.

\subsubsection{Construction of the partial waves}

The partial wave expansion for the $1\to 1$ scattering amplitude of waves with icoming momentum $k_1$ and outgoing momentum $k_2$ from a compact object takes the form
\begin{equation}
\mathcal{M}(k_1\rightarrow k_2)=(2\omega)^{3-D}\sum_\rho n_\rho^{(D)}\sum_{ij} (a_\rho(\omega))_{ji} \pi^{ij}_\rho(\cos\theta)\,,\label{eq: partial wave expansion}
\end{equation}
where $\rho$ runs over the irreducible representation of SO$(D-1)$, and $(i,j)$ labels the independent structures in the same representation. The normalization factor $n^{(D)}_\rho$ can be found in Appendix \ref{app: partial wave}. 

From the perspective of scattering amplitudes, the partial waves are functions of momenta and polarization vectors that transform in the representation $\rho$ of the SO$(D-1)$ group, which preserves the velocity vector $u^\mu$. Although the choice of $u^\mu$ seems to naively break Lorentz invariance (in a way similar to heavy quark EFT \cite{Neubert:2005mu}) we can still follow the partial wave construction in standard QFTs to prepare the three-point vertices $v^{ia}(n,e)$ that describe the minimal interactions between the worldline, the probing particles, and a massive higher-spin state
\be
n^\mu = \fft{p^\mu+u^\mu\, p\cdot u}{p\cdot u}\,,\quad e^\mu=\epsilon_\mu-p^\mu \fft{e\cdot u}{p\cdot u}\,,\label{eq: def n and e}
\ee
where $n^2=1, n\cdot e=e^2=0$. We can understand the validity of this construction by viewing the worldline as a heavy particle with momentum $P^\mu \sim m u^\mu$. The partial waves are constructed by gluing together two vertices
\begin{equation}
\pi_\rho^{ij}=v^{ia}(n_1,e_1) g_{ab}v^{jb}(n_2,e_2)\,,
\end{equation}
where $g_{ab}$ denotes the SO$(D-1)$-invariant metric on the representation $\rho$, the meaning of which will become clear soon. As detailed in Appendix \ref{app: partial wave}, our conventional normalization is
\begin{equation}
\sum_{e} v^{ia}(n,e) g_{ab}v^{jb}(n,e) =\delta^{ij}\,.
\end{equation}
Notice that the summation over the polarizations are performed by
\begin{align}
&\text{photon:}\,\qquad e^{\ast\mu}e^{\nu}\rightarrow \hat{g}^{\mu\nu}=\delta^{\mu\nu}-n^\mu n^\nu\,,\quad\nonumber\\
& \text{graviton:}\,\qquad e^{\ast \mu}e^{\ast \nu}e^\rho e^\sigma\rightarrow \frac{1}{2}\big(\hat{g}^{\mu\rho}\hat{g}^{\nu \sigma}+\hat{g}^{\nu\rho}\hat{g}^{\mu \sigma}\big)-\frac{1}{D-2}\hat{g}^{\mu\nu}\hat{g}^{\rho\sigma}\,.
\end{align}

Let us be more precise about the vertices for scalar, photon and graviton in general dimensions, and the partial waves they give. We consider general $D$ here not only because we are using dimensional regularization, but also for the convenience of future generalizations of tidal matching to higher dimensions. The special case $D=4$ can be obtained by reducing the general-dimensional constructions, effectively by keeping only the traceless-symmetric irreducible representation (i.e., the one-row tableau).
\begin{itemize}
\item Scalar 

For scalar, such vertex is unique and simple and it is labeled by the traceless symmetric representation with spin $\ell$, usually labeled by $(\ell)$
\begin{equation}
v(n)=k_\ell  \myyoung{n\tinydots n}=k_\ell (n\cdot w)^\ell\,,\quad k_\ell=\left(\frac{\big(D-3\big)_\ell}{2^\ell \big(\frac{D-3}{2}\big)_\ell}\right)^{-\frac{1}{2}}\,,
\end{equation}
where we introduced the polarization vectors $w_i\in \mathbb{C}$ to keep the vertices index free. The polarization vectors $w_i$ contracts the indices of the ith raw boxes in the Young tableau, and the traceless and symmetry properties of a given irreducible representation are guaranteed by taking them to be orthogonal and constructed modulo gauge redundancies \cite{Costa:2016hju}
\begin{equation}
w_i^2=w_i\cdot w_j=0\,,\quad w_j\sim w_j+\mathbb{C}\, w_i\,,\quad j>i\,.
\end{equation}
The details of this formalism is reviewed in Appendix \ref{app: partial wave}. 

\item Photon

For photon in general dimensions, we have two independent irreducible representations with spin $\ell$, labeled by $(\ell)$ and $(\ell,1)$
\begin{align}
 v(n,e)^{(\ell)}&=k_\ell\Big(\ft{\ell}{\ell+D-3}\Big)^{\ft{1}{2}}\myyoung{en\tinydots n} 
 \,,\nonumber\\
 v(n,e)^{(\ell,1)}& =k_\ell \Big(\ft{\ell(\ell+D-4)}{(\ell+D-3)(\ell+1) (D-3)}\Big)^{\frac{1}{2}} \myyoung{en\tinydots n,n}\,.
\end{align}

\item Graviton

We have three irreducible representations for graviton case with spin $\ell$, they are $(\ell), (\ell,1), (\ell,2)$.
\begin{align}
v(n,e)^{(\ell)}& =k_\ell \Big(\ft{(D-2)\ell(\ell-1)}{(D-3)(\ell+D-3)(\ell+D-2)}\Big)^{\frac{1}{2}}\myyoung{een\tinydots n}
\,,\nonumber\\
 v(n,e)^{(\ell,1)} & = k_{\ell} \Big(\ft{2\ell(\ell-1)(\ell+D-4)}{(D-3)(\ell+D-3)(\ell+D-2)(\ell+1)}\Big)^{\frac{1}{2}}\myyoung{een\tinydots n,n}\,,\nonumber\\
 v(n,e)^{(\ell,2)} & = k_{\ell} \Big(\ft{2(\ell-1)(\ell+D-4)}{(D-4)(D-1)(\ell+1)(\ell+D-2)}\Big)^{\ft{1}{2}} \myyoung{een\tinydots n,nn}\,.
\end{align}

\end{itemize}

To compute the partial waves, we glue two vertices. The gluing procedure is standard for the traceless symmetric rep. The building block is
\begin{equation}
\Big(\myyoung{n\tinydots n},\myyoung{\np\tinydots\np}\Big)=\big(n^{\mu_1}\cdots n^{\mu_\ell}-{\rm traces}\big)\big(\np_{\mu_1}\cdots \np_{\mu_\ell}-{\rm traces}\big)=\frac{\big(D-3)_\ell}{2^\ell \big(\frac{D-3}{2}\big)_\ell} P_\ell(n\cdot n')\,,
\end{equation}
where $P_\ell$ is the Legendre polynomials in general dimensions
\begin{equation}
P_\ell(x)=\,_2F_1\Big(-\ell,\ell+D-3,\frac{D-2}{2},\frac{1-z}{2}\Big)\,.
\end{equation}
Then, the pairing of the traceless symmetric reps can be computed by acting derivatives on the Legendre, for example
\begin{equation}
\Big(\myyoung{abn\tinydots n},\myyoung{efn\tinydots\np}\Big)=\frac{1}{\ell^2(\ell-1)^2}(a\cdot \partial_n) (b\cdot \partial_n) (e\cdot \partial_\np) (f\cdot \partial_\np) \Big(\myyoung{n\tinydots n},\myyoung{n\tinydots\np}\Big)\,.
\end{equation}

For mixed symmetric representations, we use the weight-shifting operators to remove rows until the representations become traceless symmetric forms, to which we then apply the gluing procedure described above \cite{Caron-Huot:2022jli}. The weight-shifting operators $\mathcal{D}^a$ is SO$(D-1)$ covariant differential operators that act on a tensor in the representation $\rho$ to shift it. For theoretical constructions from group representation theory, see \cite{Buric:2023ykg}. In this paper, we follow \cite{Caron-Huot:2022jli} to use the following weight-shifting operators \cite{Karateev:2018oml}
\begin{equation}
\mathcal{D}^{(h)\mu_0} =
 \left(\delta^{\mu_0}_{\mu_1} - \frac{w_1^{\mu_0}}{N_1^{(h)}} \frac{\partial}{\partial w_1^{\mu_1}} \right)
 \left(\delta^{\mu_1}_{\mu_2} - \frac{w_2^{\mu_1}}{N_2^{(h)}}\frac{\partial}{\partial w_2^{\mu_2}}  \right)\cdots
\left(\delta^{\mu_{h-1}}_{\mu_{h}} - \frac{w_h^{\mu_{h-1}}}{N_h^{(h)}{-}1}\frac{\partial}{\partial w_h^{\mu_{h}}}  \right)
\frac{\partial}{\partial w_{h\mu_{h}}}\,,
\label{todorov}
\end{equation}
where $N_i^{(h)}=D-2+m_i+m_h-i-h$. This weight-shifting operator does the job of removing other rows in the young tableau
\begin{equation}
\mathcal{D}^{(h)\mu}(m_1,m_2,\cdots m_h)\propto (m_1,\cdots m_{h-1})\,.
\end{equation}
Practically, we use the weight-shifting operators to restore the indices of antisymmetric part and then contract those indices in the pairing. For example
\begin{equation}
\Big(\myyoung{a\tinydots,b},\myyoung{c\tinydots,d}\Big)=\Big(\myyoung{a\tinydots},b\cdot\mathcal{D}^{(2)}\myyoung{c\tinydots,d}\Big)+\big(a\leftrightarrow b \big)=\Big(\mathcal{D}^{(2)\mu}\myyoung{a\tinydots,b},\mathcal{D}^{(2)}_\mu\myyoung{c\tinydots,d}\Big)\,.
\end{equation}

Applying this procedure to photon, we find the partial waves as the following
\begin{align}
 \pi_\ell^{(\ell)}(z)&=\frac{-\big(P_\ell'(z)+(z+1)P_\ell''(z)\big) H_{12}+\big(P_\ell'(z)+(z-1)P_\ell''(x)\big) V_{12}}{4\ell(\ell+D-3)} \,,\nonumber\\
 \pi_\ell^{(\ell,1)}(z)&=\frac{\big((D-3)P_\ell'(z)+(z+1)P_\ell''(z)\big) H_{12}+\big((D-3)P_\ell'(z)+(z-1)P_\ell''(z)\big) V_{12}  }{4\ell(\ell+D-3)(D-3)} \,.
\end{align}
The partial waves of graviton case are more involved, and we record them in the ancillary file and refer the readers to Appendix \ref{app: anc} for the guidance of the ancillary file. We verify the orthonormal properties of partial waves in Appendix \ref{app: partial wave}. 


\subsubsection{Partial wave coefficients from inversion formulas}

The response functions, such as those from black hole perturbation theory, are conventionally presented as partial-wave coefficients 
\begin{equation}
   i a_\rho(\omega)=\eta_\rho e^{2i\delta_\rho}-1\,,
\end{equation} so as to capture the details of multipoles. We now present the generalities used to extract the partial-wave coefficients.

The standard procedure is to start with the partial wave expansions \eqref{eq: partial wave expansion} and then use the orthonormal conditions \eqref{eq: normorth partial wave} to extract $a_\rho$. For example, for the scalar case, we have
\be
\eta_\ell e^{2i\delta_{\ell}} -1 = \frac{i\omega}{2\pi} \frac{\left(4 \pi/\omega^2\right)^{\frac{4-D}{2}}}{ 2\Gamma(\frac{D-2}{2})} \int_{-1}^1 d z\left(1-z^2\right)^{\frac{D-4}{2}} P_\ell^{(D)} (z) \mathcal{M}(\omega, z)\,.\label{eq: transform scalar gene}
\ee
As we will see later, the worldline amplitudes generally contain both infrared divergences and singularities in the forward limit, which are regularized by working in $D=4-2\epsilon_{\rm IR}$ dimensions. As shown in \cite{Ivanov:2024sds}, the partial wave transform of power-law forward singularities can be obtained for general $\ell$ using the following inversion formula
\be
\frac{\left(4 \pi \mu_{\rm IR}^2\right)^{\epsilon_{\rm IR}}}{2 \Gamma(1-\epsilon_{\rm IR})} \int_{-1}^1 d z\left(1-z^2\right)^{-\epsilon_{\rm IR}} P_\ell^{(D)} (z) \left(\frac{1-z}{2}\right)^p   =  (\pi\mu_{\rm IR}^2)^{\epsilon_{\rm IR}} \frac{\Gamma(1-\epsilon_{\rm IR} + p)\Gamma(\ell-p)}{\Gamma(-p)\Gamma(2-2\epsilon_{\rm IR} + \ell + p)}\,.\label{eq: transform formula gene}
\ee
For logarithmic contributions $\log(1-z)$, we should understand them as $(1-z)^{\# \epsilon_{\rm IR}}$ and take $\epsilon_{\rm IR}\rightarrow 0$ after performing the partial wave transformation, where the details of $\#$ depend on the amplitude solved by differential equations in section \ref{subsec: differential}. We will nevertheless use the exponential representation to remove the IR divergences (i.e., exponentiate them) to get rid of this subtlety.

It is worth noting that we should first perform \eqref{eq: transform formula gene} and then perform the analytical continuation of $\ell$ \cite{Kol:2011vg,Charalambous:2021kcz,Creci:2021rkz,Ivanov:2022qqt,Bautista:2023sdf}, which is only valid for sufficiently large $\ell$. Equivalently, as a short cut, we can also use the Froissart–Gribov formula~\cite{gribov2002gauge} to directly compute partial wave coefficients from cuts in $t=2\omega^2(z-1)$ that are proven to be analytic in spin for $\ell \geq \ell_0$.
\be
\eta_\ell e^{2i\delta_{\ell}} -1 = \frac{i\omega}{\pi^2} \frac{\left(4 \pi/\omega^2\right)^{\frac{4-D}{2}}}{ 2\Gamma(\frac{D-2}{2})} \int_{1}^\infty d z\left(z^2-1\right)^{\frac{D-4}{2}}Q_\ell^{(D)} (z) {\rm Disc}_t \mathcal{M}(\omega, z)\,, \label{eq:partl}
\ee
where $Q_\ell^{(D}(z)$ is the Gegenbauer-Q function
\be
\label{eq:FG}
Q_{\ell}^{(D)}(z)=\frac{\sqrt{\pi}\Gamma(\ell+1)\Gamma(\frac{D-2}{2})}{2^{\ell+1}\Gamma(\ell+\frac{D-1}{2})}\frac{1}{z^{\ell+D-3}} \,_2F_1\Big(\frac{\ell+D-3}{2},\frac{\ell+D-2}{2},\ell+\frac{D-1}{2},\frac{1}{z^2}\Big)\,.
\ee
Essentially, this restriction arises from the contour deformation of the amplitudes in the complex $t$-plane when deriving the Froissart–Gribov formula, because it is necessary for the integrand at complex infinity to decay sufficiently fast so that it can be dropped, given $Q_\ell^{(D)}\big|_{z\rightarrow\infty}\sim 1/z^{\ell+D-3}$ (see e.g. \cite{Correia:2020xtr}). Therefore, the restriction $\ell \geq \ell_0$ arises from the underlying Regge limit of the amplitudes in the complex $t$-plane, characterized by $\lim_{|t|\rightarrow\infty}|\mathcal{M}|<|t|^{\ell_0}$, which is, however, not well explored in worldline and black hole scattering. We emphasize that, for the same reason, the Froissart–Gribov formula produces spurious poles for physical $\ell \in \mathbb{N}$ at $\mathcal{O}(G^{2\ell+3})$. These poles are analogous to the spurious poles encountered in GR calculations of response functions \cite{Kol:2011vg,Charalambous:2021mea}, as well as the perturbative anomalous dimensions in conformal field theories, which essentially invalidate the formula for low spin around those singularities. Whenever we encounter such singularities, we must simply perform the partial wave transformation for fixed integer $\ell$ \eqref{eq: transform scalar gene}.

For the photon and graviton, the polarization structures increase the technical difficulties, including the polarization summation in \eqref{eq: partial wave expansion}, the generalization of the Froissart–Gribov formula to the spinning case\footnote{To our knowledge, we did not find any explicit spinning Froissart–Gribov formula. However, generalizations to CFTs have been constructed for generic spin using light-ray operators \cite{Kravchuk:2018htv}.}, and the appearance of higher derivatives acting on the Legendre polynomials. The polarization local basis \eqref{eq: basis A} and \eqref{eq: basis h} then play essential roles to provide simplifications. More precisely, we work out the contractions of local basis by summing over polarizations, which give us contractions matrices $M_{ee^\ast}$
\be
M_{ee^\ast}^{ij}=\sum_{ee^\ast}{\rm Poly}^i(e) {\rm Poly}^i(e^\ast)\,.
\ee
Then we use \eqref{eq: partial wave expansion}, and schematically we have
\be
a_\rho\sim i \omega^{D-3} \int dz \sum_{ij}A^{(i)} \pi_\rho^{(j)}(z) M_{ee^\ast}^{ij} (1-z^2)^{\fft{D-4}{2}}\,.
\ee
To deal with terms $d^n/dz^n P_\ell(z)$ in this integral, we consider integration by parts so as to reduce the problem to a scalar inversion. We can then use \eqref{eq: transform formula gene} or rewrite the resulting scalar inversion in terms of the Froissart–Gribov formula \eqref{eq:partl}\footnote{A similar trick was used in \cite{Caron-Huot:2021kjy} to compute spinning OPE data in holographic CFTs.}. We then insert $a_\rho$ back into the partial wave expansions \eqref{eq: partial wave expansion} and note that taking $D=4-2\epsilon$ with $\epsilon\rightarrow 0$ reduces partial waves in general dimensions to those in $D=4$ (which only contain the $(\ell)$ irrep). We can then collect the phase shifts with polarization structure $a_{(\ell)}^{(i)} = a_{(\ell)} {\rm Poly}^{(i)}$.

\subsection{Retarded versus in-out amplitude}
\label{subsec: retarded vs Feynman}

We first discuss the relation between the retarded amplitude and in-out amplitude. In the standard quantum field theory, the scattering amplitude is defined to be 
\begin{equation}
    \langle {\rm BH}| a_{\rm out} (\boldsymbol{k}_2) a^\dagger_{\rm in} (\boldsymbol{k}_1) | {\rm BH} \rangle = 2\pi  \delta(k_2^0 - k_1^0) 2k_1^0 S_F(\boldsymbol{k}_2, \boldsymbol{k}_1) ~,
\end{equation}
which can also be obtained from the LSZ reduction on the time-ordered correlation function. In this type of observables, the relevant propagator is the Feynman propagator (time-ordered Green's function), which is, for the scalar example, given by
\begin{equation}
    G_{\rm F} = \frac{i}{p_0^2 - \boldsymbol{p}^2 + i0} ~.
\end{equation}
However, the classical observables are so-called \textit{retarded} amplitude, which is defined to be
\begin{equation}
    \langle {\rm BH}|[a_{\rm out}(\boldsymbol{k}_2), a_{\rm in}^\dagger(\boldsymbol{k}_1)]|{\rm BH}\rangle = 2 \pi \delta (k_2^0 - k_1^0) 2 k_1^0 S_R(\boldsymbol{k}_2,\boldsymbol{k}_1) ~.
\end{equation}
This amplitude can also be obtained from the LSZ reduction on the retarded correlation function \cite{Caron-Huot:2023vxl}. In this type of observables, one should use the retarded propagator
\begin{equation}
    G_{\rm R} = \frac{i}{(p_0+i0)^2 -\boldsymbol{p}^2} ~.
\end{equation}
According to the identity
\begin{equation}
\begin{aligned}
    \frac{1}{p_0^2 - \boldsymbol{p}^2 + i0} & = {\rm PV} \frac{1}{p_0^2 - \boldsymbol{p}^2} - i \pi \delta(p_0^2 -\boldsymbol{p}^2)  ~, \\
    \frac{1}{(p_0 + i0)^2 -\boldsymbol{p}^2} & = \frac{1}{p_0^2 - \boldsymbol{p}^2 +  {\rm sign}(p_0) i0} = {\rm PV} \frac{1}{p_0^2 -\boldsymbol{p}^2} - i \pi {\rm sign}(p_0) \delta(p_0^2 - \boldsymbol{p}^2) ~,
\end{aligned}
\end{equation}
we notice that when $p_0>0$,
\begin{equation}
    G_{\rm F} = G_{\rm R} ~.
\end{equation}
For $p_0<0$, we instead have the relation
\begin{equation}
    G_{\rm R} = G_F - 2\pi \delta(p_0^2 - \boldsymbol{p}^2) ~.
\end{equation}
The above relation can be combined into unified equation
\begin{equation}
\label{eq:retarded_feynman relation}
    G_R = G_F - G_< ~,
\end{equation}
where $G_< = 2\pi \delta(p_0^2 - \boldsymbol{p}^2)$ is the negative frequency Wightman function. In the physical kinematics, the incoming particle has positive frequency. In this regime, it is straightforward to see that $p_0>0$ is satisfied and therefore there is no difference between the retarded and the Feymann propagator.
In the following, we will use the Feymann Green's function for the calculation in the conservative sector. Later, in Section~\ref{subsec:exp rep}, we are going to see that the Feynman prescription gives us a lot of simplification as this allows as to use the perturbative unitarity of the S-matrix. However, the payoff is that the results are not good playgrounds if one is interested in the analyticity in $p_0$ (see \cite{Correia:2025enx} for the relevant discussion on analyticity in $p_0$)

The above analysis can be easily extended to the case of photon ($\gamma$), graviton ($h$) and recoil propagators ($\rm wl$) (see Section~\ref{subsec:recoil} for more detailed discussion) with
\begin{equation}
    G_F^\gamma = \frac{i}{p_0^2 - \boldsymbol{p}^2 + i0} \eta_{\mu\nu} ~, \quad G_F^h = \frac{i}{p_0^2 - \boldsymbol{p}^2 + i0} \Pi_{\mu\nu\rho \sigma} ~, \quad G_F^{\rm wl} = \frac{-i}{\omega^2 + i0} ~,
\end{equation}
where 
\begin{equation}
    \Pi_{\mu\nu\rho\sigma} = \frac{1}{2} \Bigg(\eta_{\mu\rho} \eta_{\nu\sigma} + \eta_{\mu \sigma} \eta_{\nu\rho} - \frac{2}{D-2} \eta_{\mu\nu} \eta_{\rho \sigma} \Bigg) ~,
\end{equation}
is the traceless-symmetric projection. 

For tidal response function, the relation in Eq.~\eqref{eq:retarded_feynman relation} still holds. If the system is in a stationary ground state, the negative frequency Wightman function only has $\omega<0$ component while positive frequency Wightman function only has support on $\omega>0$. In this case, we can identify $G_F$ with $G_R$ for the tidal response function in $\omega>0$ region. For Schwarzschild black hole, this condition is satisfied for the Boulware state. However, this condition is no longer true if the system is in excited states or thermal states. We provide more detailed discussions on this point in Appendix.~\ref{app:wightman_func}.

From now on, unless otherwise specified, we restrict to the positive-frequency sector, where the retarded and Feynman propagators coincide. As a result, in–out amplitudes and in–in response observables are identical, and we will not distinguish between them. 

\subsection{Exponential Representation of S-Matrix}
\label{subsec:exp rep}

Being able to restrict our attention to the in-out amplitude, allows us to use the properties of the S-matrix (which time-evolves in to out states) to simplify our calculations.

The standard parametrization of the S-matrix is given by 
\begin{equation}
    S = 1 + i T ~,
\end{equation}
where $T$ is the scattering operator. In the context of worldline EFTs, we consider the following 1-to-1 massless particle scattering with the worldline or compact object
\begin{equation}
    \langle k_2, h_2| T |k_1, h_1 \rangle =  \mathcal{M}(k_1,h_1 \rightarrow k_2,h_2) (2\pi) \delta(u \cdot (k_1 + k_2)) ~.
\end{equation}
The $\delta$ function makes manifest energy conservation due to the time translation invariance along the worldline. Nevertheless, in order to make manifest unitarity of the $S$-matrix
\begin{equation}
    S^\dagger S = \mathbbm{1} ~,
\end{equation}
we may consider another natural parametrization of the S-matrix
\begin{equation}
\label{eq:exp_rep}
    S = \exp (i \Delta) ~,
\end{equation}
with
\begin{equation}
    \langle k_2, h_2| \Delta |k_1, h_1 \rangle = i \Delta (k_1,h_1 \rightarrow k_2,h_2) (2\pi) \delta(u \cdot (k_1 + k_2)) ~,
\end{equation}
which is known as the exponential representation \cite{Damgaard:2023ttc}.
In the calculation of Raman scattering amplitude, the scattering matrix can be perturbatively computed in terms of Newton's constant $G$. Formally, we can write the perturbative series as
\begin{equation}
\begin{aligned}
    \Delta & = G \Delta_G + G^2 \Delta_{G^2} + G^3 \Delta_{G^3} + G^4 \Delta_{G^4}  \\
    T & = G T_G + G^2 T_{G^2} + G^3 T_{G^3} + G^4 T_{G^4} ~,
\end{aligned}
\end{equation}
Once plugging the above perturbative expression into the exponential parameterization in Eq.~\eqref{eq:exp_rep}, we obtain the relation between the $\Delta$ and $T$ matrix
\begin{equation}
\begin{aligned}
\label{eq:reverseU1}
    \Delta_G & = T_G ~, \\
    \Delta_{G^2} & = T_{G^2} - \frac{i}{2} T_{G} T_{G}~,  \\
    \Delta_{G^3} & = T_{G^3}- \frac{i}{2} \Bigg(T_{G^1} T_{G^2} + T_{G^2} T_{G^1}\Bigg) - \frac{1}{3} T_{G} T_{G} T_{G} ~.
\end{aligned}
\end{equation}
This expression can be even made simplier by making use of the perturbative unitarity
\begin{equation}
    T-T^{\dagger}=i T T^{\dagger}  \Rightarrow 2 {\rm Im} T_{G^2} = T_{G} T^\dagger_{G} = T_{G} T_{G} , ~ 2 {\rm Im} T_{G^3} = T_{G^2} T_{G}^\dagger + T_{G} T_{G^2}^\dagger ~,
\end{equation}
which naturally generates the following simple formulas
\begin{equation}
        \begin{aligned}
            i \Delta_G & = i T_{G} ~, \\
            i  \Delta_{G^2} & = {\rm Im} (i T_{G^2}) \\
            i \Delta_{G^3} & = {\rm Im}(i T_{G^3}) - \frac{1}{6} (i T_{G}) (i T_{G}) (i T_{G}) ~.
        \end{aligned}
\end{equation}
In the practical calculation, we shall further use the so-called ``reverse unitarity" by replacing the on-shell delta functions that appear in the phase space integral to the difference of propagators with opposite $i0$ prescription
\begin{equation}
    (-1)^n \frac{i 2 \pi}{n !} \delta^{(n)}\left(p^2\right)=\frac{1}{\left(p^2-i 0\right)^{n+1}}-\frac{1}{\left(p^2+i 0\right)^{n+1}} ~.
\end{equation}
For example, the evaluation of $\Delta_{G^3}$ can be done by directly computing the imaginary part of the $G^3$ amplitude and subtract the iterated one-particle cuts
\begin{equation}
    \begin{gathered}
    \begin{tikzpicture}[line width=1,photon/.style={decorate, decoration={snake, amplitude=1pt, segment length=6pt}}]
    \draw[line width = 1, photon] (0,0) -- (1,1);
    \draw[line width = 1, photon] (0,0) -- (1,-1);
    \filldraw[fill=gray!40, line width=1.2,yshift=2](0,0) circle (0.6) node {\small $\Delta_{G^3}$};
    \end{tikzpicture}
    \end{gathered}
    = {\rm Im} \left[     
    \begin{gathered}
    \begin{tikzpicture}[line width=1,photon/.style={decorate, decoration={snake, amplitude=1pt, segment length=6pt}}]
    \draw[line width = 1, photon] (0,0) -- (1,1);
    \draw[line width = 1, photon] (0,0) -- (1,-1);
    \filldraw[fill=gray!40, line width=1.2,yshift=2](0,0) circle (0.6) node {\small $T_{G^3}$};
    \end{tikzpicture} 
    \end{gathered}
    \right]
    - \frac{1}{6} ~ 
    \begin{gathered}
    \begin{tikzpicture}[line width=1,photon/.style={decorate, decoration={snake, amplitude=1pt, segment length=6pt}}]
    \draw[line width = 1, photon] (0,-1.0) -- (0,0.0);
    \draw[line width = 1, photon] (0,-1.0) -- (1,-1.5);
    \draw[line width = 1, photon] (0,0.0) -- (0,1.0);
    \draw[line width = 1, photon] (0,1.0) -- (1,1.5);
    \draw[dashed,Orange] (-0.5,0.6) -- (0.5,0.6);
    \draw[dashed,Orange] (-0.5,-0.45) -- (0.5,-0.45);
    \filldraw[fill=gray!40, line width=1.2,yshift=2](0,-1.0) circle (0.3) node {\small $T_{G}$};
    \filldraw[fill=gray!40, line width=1.2,yshift=2](0,0.0) circle (0.3) node {\small $T_{G}$};
    \filldraw[fill=gray!40, line width=1.2,yshift=2](0,1.0) circle (0.3) node {\small $T_{G}$};
    \end{tikzpicture} 
    \end{gathered}
    ~.
\end{equation}

Particularly in the gravitational Raman scatterings, exponential representation greatly simplifies the calculation. On one hand, the Raman amplitude has universal IR divergence due the long-range Newtonian potential. At the $L$-loop order, it will generate $1/\epsilon_{\rm IR}^L$ divergence in the dimensional regularization and will exponentiate in the end. Therefore, in the exponential representation, the universal IR phase will only show up in the tree-level $\Delta_{G}$. All the loop-level $\Delta$-matrix will be IR safe. On the other hand, since we have already put the scattering matrix on the exponent, we can directly extract phase shifts $\delta_{G^i}, i = 1,2,3 \cdots$ order by the partial wave transforming the $\Delta_{G^i}$ matrix, without a need to perform further subtractions.

\section{The Scattering Amplitude in EFT}
\label{sec:ampeft}
In this section, we discuss the computation of worldline EFT amplitudes. In particular, we simplify the Feynman rules by using the background field method via the insertion of the asymptotic metric sourced by the worldline. In the graviton case, this also requires including recoil to ensure gauge invariance. We then apply modern amplitude techniques, such as integration-by-parts (IBP) \cite{Chetyrkin:1981qh} and differential equations \cite{Henn:2013pwa,Henn:2014qga}, to compute loop diagrams and extract the response functions in the partial wave basis, and present our results.

\subsection{Background field action, recoil operator and Feynman diagrams}
\label{subsec:recoil}

We start the discussion by reviewing the background field method and illustrate the recoil operator that is needed for the calculation of gravitational perturbations \cite{Cheung:2023lnj,Cheung:2024byb}. We take the Schwarzschild background within the isotropic coordinate \cite{Duff:1973zz} 
\begin{equation}
    \bar g_{\mu\nu} = \left(1+\frac{\alpha}{4|\boldsymbol{r}|^{D-3}}\right)^{4 /(D-3)}\left(\eta_{\mu \nu}+u_\mu u_\nu\right)-\left(\frac{1-\frac{\alpha}{4 | \boldsymbol{r}|^{D-3}}}{1+\frac{\alpha}{4|\boldsymbol{r}|^{D-3}}}\right)^2 u_\mu u_\nu ~,
\end{equation}
where 
\begin{equation}
    \alpha = 2 G M n_d \mu^{4-d} ~, \quad n_d \equiv \frac{4 \pi^{\frac{3-d}{2}} \Gamma\left(\frac{d-1}{2}\right)}{d-2} ~.
\end{equation}
This background can be perturbatively expanded into the form as
\begin{equation}
    \begin{aligned}
\bar{g}_{\alpha \beta} & =\eta_{\alpha \beta}-u_\alpha u_\beta\left(-\frac{\alpha}{r^{D-3}}+\frac{1}{2} \frac{\alpha^2}{r^{2(D-3)}}\right) \\
& +\left(\eta_{\alpha \beta}+u_\alpha u_\beta\right)\left(\frac{1}{D-3} \frac{\alpha}{r^{D-3}}-\frac{D-7}{8(D-3)^2} \frac{\alpha^2}{r^{2(D-3)}}\right) \\
& +\cdots ~,
\end{aligned}
\end{equation}
which aligns with the order by order expansion in $G$. On the top of this background, we can add scalar, photon and graviton perturbations, which can be written as
\begin{equation}
    S = - \frac{1}{2} \int d^D x \sqrt{-g}g^{\mu\nu} \partial_\mu \phi \partial_\nu \phi = - \frac{1}{2} \int d^D x (\sqrt{-g} g^{\mu\nu} - \eta^{\mu\nu}) \partial_\mu \phi \partial_\nu \phi - \frac{1}{2} \int d^D x \eta^{\mu\nu} \partial_\mu \phi \partial_\nu \phi ~,
\end{equation}
where we treat the last term as the free field theory in flat space and the first term as perturbations. For the electromagnetic perturbations, we have
\begin{equation}
    S_\gamma=-\frac{1}{4} \int d^Dx\sqrt{-\bar g}F_{\mu\nu}^2\, ~.
\end{equation}
To fix the gauge, we choose a background Feynman gauge-fixing term
\begin{equation}
    S_{\gamma,\rm gf}=-\frac{1}{2} \int d^Dx\sqrt{-\bar g} (\bar \nabla_\mu A^\mu)^2\, ~.
\end{equation}
Then, the total action becomes
\begin{equation}
    S_\gamma + S_{\gamma,{\rm gf}} =  -\frac{1}{2} \int d^Dx\sqrt{-\bar g} [ (\bar\nabla_\mu A_\nu)^2 + \bar R^{\mu\nu} A_\mu A_\nu] ~.
\end{equation}

The gravitational perturbations are more complicated. We start from the Einstein-Hilbert action
\begin{equation}
    S_h = \frac{1}{16\pi G} \int d^D x \sqrt{-g} R ~,
\end{equation}
with the gauge fixing term
\begin{equation}
    S_{h,{\rm gf}} =  \int d^Dx \sqrt{-\bar g} G_{\mu} G^\mu
\end{equation}
where
\begin{equation}
    G_\mu = \nabla^\nu \delta g_{\mu\nu} - \frac{1}{2} \nabla_\mu \delta g ~, \quad \delta g \equiv \delta g_{\mu\nu} \bar g^{\mu\nu} ~.
\end{equation}
Now, we can directly plug in the perturbative expansion
\begin{equation}
    g_{\mu\nu} = \bar g_{\mu\nu} + \kappa \delta g_{\mu\nu} ~,
\end{equation}
where $\kappa \equiv \sqrt{32 \pi G}$ to get the background field action
\begin{equation}
\begin{aligned}
    S_h + S_{h,{\rm gf}}& = \int d^D x \sqrt{-\bar g} \Bigg[ \frac{1}{2} \delta g^{\alpha \beta} \bar \nabla^2 \delta g_{\alpha \beta} - \frac{1}{4} \delta g \bar \nabla^2 \delta g +  \delta g^{\alpha\beta} \delta g^{\gamma \mu} \bar R_{\alpha\gamma\beta\mu} \\
    & \quad + \Bigg( \delta g^\gamma{}_\alpha \delta g_{\beta \gamma} - \delta g_{\alpha\beta} \delta g \Bigg) \bar R^{\alpha \beta} + \Bigg(- \frac{1}{2} \delta g_{\alpha\beta} \delta g^{\alpha \beta} + \frac{1}{4} \delta g^2 \Bigg) \bar R \Bigg] ~.
\end{aligned}
\end{equation}
At the first glance, this is all we need to perturbation the calculations for gravitational perturbations. However, from the Polyakov form the worldline action, we notice that the metric has direct coupling with the worldline. Therefore, when we perform the perturbation on the worldline coordinate $x^\mu = \bar x^\mu + \delta x^\mu = u^\mu \tau + \delta x^\mu$, we shall get
\begin{equation}
\begin{aligned}
M \int d \tau\left[-\frac{1}{2}+\frac{1}{2} \dot{x}^\mu \dot{x}^\nu g_{\mu \nu}\left(x\right)\right]=M \int & d \tau\left[-\frac{1}{2}+\frac{1}{2} \dot{\bar{x}}^2-\delta x \ddot{\bar{x}}-\delta x^\rho \dot{\bar{x}}^\mu \dot{\bar{x}}^\nu \bar{\Gamma}_{\rho \mu \nu}\left(\bar{x}\right)+\frac{\kappa}{2} \dot{\bar{x}}^\mu \dot{\bar{x}}^\nu \delta g_{\mu \nu}\left(\bar{x}\right) \right. \\
& +\frac{1}{2} \dot{\bar{x}}^\mu \dot{\bar{x}}^\nu \bar{\nabla}_\mu \delta x^\rho \bar{\nabla}_\nu \delta x_{\rho}+\frac{1}{2} \delta x^\rho \delta x^\sigma \dot{\bar{x}}^\mu \dot{\bar{x}}^\nu \bar{R}_{\nu \rho \sigma \mu}\left(\bar{x}\right) \\
& \left.-\kappa\delta x_{\rho} \dot{\bar{x}}^\mu \dot{\bar{x}}^\nu \delta \Gamma_{\mu \nu}^\rho\left(\bar{x}\right)+\cdots\right] ~.
\end{aligned}
\end{equation}
The terms $\delta x \ddot{\bar{x}}, \dot{\bar x}^\mu \dot{\bar x}^\nu \delta g_{\mu\nu}$ vanish due to the background equation of motion \footnote{The linear in $\delta g$ part vanish because the prefactor sources the energy momentum tensor for the background Einstein equation.}. The terms $\delta x^\rho \dot{\bar x}^\mu \dot{\bar x}^\nu \bar{\Gamma}_{\rho \mu \nu}, \delta x^\rho \delta x^\sigma  \dot{\bar x}^\mu \dot{\bar x}^\nu \bar{R}_{\nu\rho\sigma\mu}$ represents the self-energy term correction due to the off-shell gravitational potential. Physically, it corresponds to the mass renormalization \cite{Cheung:2023lnj,Cheung:2024byb}. Technically, it only generates scaleless integral which vanish in dimensional regularization. Collecting the rest of the terms, we get
\begin{equation}
    S_{\rm recoil} = M \int d\tau \Bigg[ \frac{1}{2} \dot{\delta x}^2 - \kappa \delta x_\rho \dot{\bar{x}}^\mu \dot{\bar x}^\nu \delta \Gamma^\rho_{\mu\nu}(\bar x) \Bigg] ~.
\end{equation}
Now, we see that $\delta x$ is essentially the gapless excitation of a free propagating mode along the worldline that can be easily integrated out to get a non-local in time effective action
\begin{equation}
    S_{\rm recoil} = (16 \pi G M) \int d \tau \delta \Gamma^\mu\left(\bar{x}\right) \frac{1}{\partial_\tau^2} \delta \Gamma_{ \mu}\left(\bar{x}\right) ~.
\end{equation}
With the effective action $S_h + S_{h,{\rm gf}} + S_{\rm recoil}$, we have all the ingredients to perform the amplitude calculation. 

In this setup, the entire scattering amplitude for massless waves against compact objects contain two parts. The first part is the scattering against the background metric sourced by the object
\begin{equation}
\begin{aligned}
\label{eq:farzone}
\hspace{-0.7cm}
\left[
\begin{gathered}
    \begin{tikzpicture}[line width=1,photon/.style={decorate, decoration={snake, amplitude=1pt, segment length=6pt}}]
    \draw[line width = 1, photon] (0,0) -- (1,1);
    \draw[line width = 1, photon] (0,0) -- (1,-1);
    \filldraw[fill=gray!5, line width=1.2,yshift=2](0,0) circle (0.6) node {\small star};
    \end{tikzpicture}
\end{gathered}
\right]_{\rm BG}
& = 
\begin{gathered}
    \begin{tikzpicture}[line width=1,photon/.style={decorate, decoration={snake, amplitude=1pt, segment length=6pt}}]
    \draw[line width = 1, photon] (0,0) -- (1,1);
    \draw[line width = 1, photon] (0,0) -- (1,-1);
    \draw[line width = 1, dashed] (-1,0) node[midway, above, xshift=-10] {\footnotesize $G M /r$} -- (0,0);
    \filldraw[fill=gray!5, line width=1.2](-1,0) circle (0.15) node {$\times$};
    \end{tikzpicture}
\end{gathered}
+
\begin{gathered}
    \begin{tikzpicture}[line width=1,photon/.style={decorate, decoration={snake, amplitude=1pt, segment length=6pt}}]
    \draw[line width = 1, photon] (0,0.5) -- (1,1.5) node[pos=1.0, below,xshift=-10,yshift=-12] {\footnotesize $k_2$};
    \draw[line width = 1, photon] (0,-0.5) -- (1,-1.5) node[ pos=1.0, above,xshift=-10,yshift=12] {\footnotesize $k_1$};
    \draw[line width = 1, photon] (0,-0.5) -- (0,0.5);
    \draw[line width = 1, dashed] (-1,-0.5) -- (0,-0.5) node[pos=0.5, below] {\footnotesize $\ell_1$};
    \draw[line width = 1, dashed] (-1,0.5) -- (0,0.5) node[pos=0.5, above] {\footnotesize $\ell_2$};
    \filldraw[fill=gray!5, line width=1.2](-1,-0.5) circle (0.15) node {$\times$};
    \filldraw[fill=gray!5, line width=1.2](-1,0.5) circle (0.15) node {$\times$};
    \end{tikzpicture}
\end{gathered}
+
\begin{gathered}
    \begin{tikzpicture}[line width=1,photon/.style={decorate, decoration={snake, amplitude=1pt, segment length=6pt}}]
    \draw[line width = 1, photon] (0,0.0) -- (1,1.0);
    \draw[line width = 1, photon] (0,0.0) -- (1,-1.0);
    \draw[line width = 1, dashed] (-1,-0.5) -- (0,0.0);
    \draw[line width = 1, dashed] (-1,0.5) -- (0,0.0);
    \filldraw[fill=gray!5, line width=1.2](-1,-0.5) circle (0.15) node {$\times$};
    \filldraw[fill=gray!5, line width=1.2](-1,0.5) circle (0.15) node {$\times$};
    \end{tikzpicture}
\end{gathered}
+
\begin{gathered}
    \begin{tikzpicture}[line width=1,photon/.style={decorate, decoration={snake, amplitude=1pt, segment length=6pt}}]
    \draw[line width = 1, photon] (0,1.0) -- (1,2.0) node[pos=1.0, below,xshift=-10,yshift=-12] {\footnotesize $k_2$};
    \draw[line width = 1, photon] (0,-1.0) -- (1,-2.0) node[ pos=1.0, above,xshift=-10,yshift=12] {\footnotesize $k_1$};
    \draw[line width = 1, photon] (0,-1.0) -- (0.0,0.0);
    \draw[line width = 1, photon] (0.0,0.0) -- (0.0,1.0) ;
    \draw[line width = 1, dashed] (-1,-1.0) -- (0,-1.0) node[pos=0.5, below] {\footnotesize $\ell_1$};
    \draw[line width = 1, dashed] (-1,1.0) -- (0,1.0) node[pos=0.5, above] {\footnotesize $\ell_2$};
    \draw[line width = 1, dashed] (-1,0.0) -- (0,0.0);
    \filldraw[fill=gray!5, line width=1.2](-1,-1.0) circle (0.15) node {$\times$};
    \filldraw[fill=gray!5, line width=1.2](-1,1.0) circle (0.15) node {$\times$};
    \filldraw[fill=gray!5, line width=1.2](-1,0.0) circle (0.15) node {$\times$};
    \end{tikzpicture}
\end{gathered}
\\
& \quad +
\begin{gathered}
    \begin{tikzpicture}[line width=1,photon/.style={decorate, decoration={snake, amplitude=1pt, segment length=6pt}}]
    \draw[line width = 1, photon] (0,0.0) -- (1,1.0);
    \draw[line width = 1, photon] (0,-1.0) -- (1,-2.0);
    \draw[line width = 1, photon] (0.0,0.0) -- (0.0,-1.0);
    \draw[line width = 1, dashed] (-1,-1.0) -- (0,-1.0);
    \draw[line width = 1, dashed] (-1,0.5) -- (0,0.0);
    \draw[line width = 1, dashed] (-1,-0.5) -- (0,0.0);
    \filldraw[fill=gray!5, line width=1.2](-1,-1.0) circle (0.15) node {$\times$};
    \filldraw[fill=gray!5, line width=1.2](-1,0.5) circle (0.15) node {$\times$};
    \filldraw[fill=gray!5, line width=1.2](-1,-0.5) circle (0.15) node {$\times$};
    \end{tikzpicture}
\end{gathered}
+
\begin{gathered}
    \begin{tikzpicture}[line width=1,photon/.style={decorate, decoration={snake, amplitude=1pt, segment length=6pt}}]
    \draw[line width = 1, photon] (0,0.0) -- (1,1.0);
    \draw[line width = 1, photon] (0,-1.0) -- (1,-2.0);
    \draw[line width = 1, photon] (0.0,0.0) -- (0.0,-1.0);
    \draw[line width = 1, dashed] (-1,-1.5) -- (0,-1.0);
    \draw[line width = 1, dashed] (-1,0.0) -- (0,0.0);
    \draw[line width = 1, dashed] (-1,-0.5) -- (0,-1.0);
    \filldraw[fill=gray!5, line width=1.2](-1,-1.5) circle (0.15) node {$\times$};
    \filldraw[fill=gray!5, line width=1.2](-1,0.0) circle (0.15) node {$\times$};
    \filldraw[fill=gray!5, line width=1.2](-1,-0.5) circle (0.15) node {$\times$};
    \end{tikzpicture}
\end{gathered}
+
\begin{gathered}
    \begin{tikzpicture}[line width=1,photon/.style={decorate, decoration={snake, amplitude=1pt, segment length=6pt}}]
    \draw[line width = 1, photon] (0,0.0) -- (1,1.0);
    \draw[line width = 1, photon] (0,0.0) -- (1,-1.0);
    \draw[line width = 1, dashed] (-1,-1.0) -- (0,0.0);
    \draw[line width = 1, dashed] (-1,1.0) -- (0,0.0);
    \draw[line width = 1, dashed] (-1,0.0) -- (0,0.0);
    \filldraw[fill=gray!5, line width=1.2](-1,-1.0) circle (0.15) node {$\times$};
    \filldraw[fill=gray!5, line width=1.2](-1,1.0) circle (0.15) node {$\times$};
    \filldraw[fill=gray!5, line width=1.2](-1,0.0) circle (0.15) node {$\times$};
    \end{tikzpicture}
\end{gathered}
+\mathcal{O}(G^4)
\\
&\quad  +
\begin{gathered}
    \begin{tikzpicture}[line width=1,photon/.style={decorate, decoration={snake, amplitude=1pt, segment length=6pt}}]
    \draw[line width = 1, photon] (-1,0) -- (0,1);
    \draw[line width = 1, photon] (-1,0) -- (0,-1);
    \filldraw[fill=gray!40, draw=black, line width=1.2]
    (-1,0) circle (0.15);
    \end{tikzpicture}
\end{gathered}
    \quad + \begin{gathered}
    \begin{tikzpicture}[line width=1,photon/.style={decorate, decoration={snake, amplitude=1pt, segment length=6pt}}]
    \draw[line width = 1, photon] (-1,0.5) -- (0,1.5);
    \draw[line width = 1, photon] (0,-0.5) -- (1,-1.5);
    \draw[line width = 1, photon] (0,-0.5) -- (-1,0.5);
    \draw[line width = 1, dashed] (-1,-0.5) -- (0,-0.5);
    \filldraw[fill=gray!5, line width=1.2](-1,-0.5) circle (0.15) node {$\times$};
    \filldraw[fill=gray!40, draw=black, line width=1.2](-1,0.5) circle (0.15);
    \end{tikzpicture}
\end{gathered}
+
\begin{gathered}
    \begin{tikzpicture}[line width=1,photon/.style={decorate, decoration={snake, amplitude=1pt, segment length=6pt}}]
    \draw[line width = 1, photon] (-1,-0.5) -- (0,-1.5);
   \draw[line width = 1, photon] (-1,-0.5) -- (1,1.5);
    \draw[line width = 1, dashed] (-1,0.5) -- (0,0.5);
    \filldraw[fill=gray!5, line width=1.2](-1,0.5) circle (0.15) node {$\times$};
    \filldraw[fill=gray!40, draw=black, line width=1.2](-1,-0.5) circle (0.15);
    \end{tikzpicture}
\end{gathered}
+ \mathcal{O}(G^2 \text{recoil})\,,
\end{aligned}
\end{equation}
which essentially describes the scattering with insertions of the mass monopole on the worldline. The third line includes the effects of recoil operators in the case of the gravitational wave scattering. The second part can be intuitively understood as the scattering against the surface of stars (or horizon degrees of freedom for black holes), which includes the tidal response and the dissipation effects
\begin{equation}
\begin{aligned}
\left[
\begin{gathered}
    \begin{tikzpicture}[line width=1,photon/.style={decorate, decoration={snake, amplitude=1pt, segment length=6pt}}]
    \draw[line width = 1, photon] (0,0) -- (1,1);
    \draw[line width = 1, photon] (0,0) -- (1,-1);
    \filldraw[fill=gray!5, line width=1.2,yshift=2](0,0) circle (0.6) node {\small star};
    \end{tikzpicture}
\end{gathered}
\right]_{\rm tides}
& =
\begin{gathered}
    \begin{tikzpicture}[line width=1,photon/.style={decorate, decoration={snake, amplitude=1pt, segment length=6pt}}]
    \draw[line width = 1, photon] (0,0.5) -- (1,1.5);
    \draw[line width = 1, photon] (0,-0.5) -- (1,-1.5);
    \draw[line width = 1, dashed, double] (0,-0.5) -- (0,0.5);
    \filldraw[fill=black, line width=1.2](0,0.5) circle (0.15) node[left]{\small$Q\;$};
    \filldraw[fill=black, line width=1.2](0,-0.5) circle (0.15) node[left]{\small$Q\;$};
    \end{tikzpicture}
\end{gathered}
+
\begin{gathered}
    \begin{tikzpicture}[line width=1,photon/.style={decorate, decoration={snake, amplitude=1pt, segment length=6pt}}]
    \draw[line width = 1, photon] (0,0.5) -- (1,1.5);
    \draw[line width = 1, photon] (0,-0.5) -- (1,-1.5);
    \draw[line width = 1, dashed, double] (0,-0.5) -- (0,0.5);
    \draw[line width = 1, dashed] (0,1.2) -- (0.5,1.0);
    \filldraw[fill=black, line width=1.2](0,0.5) circle (0.15);
    \filldraw[fill=black, line width=1.2](0,-0.5) circle (0.15);
    \filldraw[fill=gray!5, line width=1.2](0,1.2) circle (0.15) node {$\times$};
    \end{tikzpicture}
\end{gathered}
+
\begin{gathered}
    \begin{tikzpicture}[line width=1,photon/.style={decorate, decoration={snake, amplitude=1pt, segment length=6pt}}]
    \draw[line width = 1, photon] (0,0.5) -- (1,1.5);
    \draw[line width = 1, photon] (0,-0.5) -- (1,-1.5);
    \draw[line width = 1, dashed, double] (0,-0.5) -- (0,0.5);
    \draw[line width = 1, dashed] (0,-1.2) -- (0.5,-1.0);
    \filldraw[fill=black, line width=1.2](0,0.5) circle (0.15);
    \filldraw[fill=black, line width=1.2](0,-0.5) circle (0.15);
    \filldraw[fill=gray!5, line width=1.2](0,-1.2) circle (0.15) node {$\times$};
    \end{tikzpicture}
\end{gathered}
+
\begin{gathered}
    \begin{tikzpicture}[line width=1,photon/.style={decorate, decoration={snake, amplitude=1pt, segment length=6pt}}]
    \draw[line width = 1, photon] (0,0.5) -- (1,1.5);
    \draw[line width = 1, photon] (0,-0.5) -- (1,-1.5);
    \draw[line width = 1, dashed, double] (0,-0.5) -- (0,0.5);
    \draw[line width = 1, dashed] (0,1.2) -- (0.5,1.0);
    \draw[line width = 1, dashed] (0,-1.2) -- (0.5,-1.0);
    \filldraw[fill=black, line width=1.2](0,0.5) circle (0.15);
    \filldraw[fill=black, line width=1.2](0,-0.5) circle (0.15);
    \filldraw[fill=gray!5, line width=1.2](0,-1.2) circle (0.15) node {$\times$};
    \filldraw[fill=gray!5, line width=1.2](0,1.2) circle (0.15) node {$\times$};
    \end{tikzpicture}
\end{gathered}
\\
& \quad +
\begin{gathered}
    \begin{tikzpicture}[line width=1,photon/.style={decorate, decoration={snake, amplitude=1pt, segment length=6pt}}]
    \draw[line width = 1, photon] (0,0.5) -- (1,1.5);
    \draw[line width = 1, photon] (0,-0.5) -- (1,-1.5);
    \draw[line width = 1, dashed, double] (0,-0.5) -- (0,0.5);
    \draw[line width = 1, dashed] (0,1.0) -- (0.5,1.0);
    \draw[line width = 1, dashed] (0,1.7) -- (0.5,1.0);
    \filldraw[fill=black, line width=1.2](0,0.5) circle (0.15);
    \filldraw[fill=black, line width=1.2](0,-0.5) circle (0.15);
    \filldraw[fill=gray!5, line width=1.2](0,1.0) circle (0.15) node {$\times$};
    \filldraw[fill=gray!5, line width=1.2](0,1.7) circle (0.15) node {$\times$};
    \end{tikzpicture}
\end{gathered}
+
\begin{gathered}
    \begin{tikzpicture}[line width=1,photon/.style={decorate, decoration={snake, amplitude=1pt, segment length=6pt}}]
    \draw[line width = 1, photon] (0,0.5) -- (1,1.5);
    \draw[line width = 1, photon] (0,-0.5) -- (1,-1.5);
    \draw[line width = 1, dashed, double] (0,-0.5) -- (0,0.5);
    \draw[line width = 1, dashed] (0,-1.0) -- (0.5,-1.0);
    \draw[line width = 1, dashed] (0,-1.7) -- (0.5,-1.0);
    \filldraw[fill=black, line width=1.2](0,0.5) circle (0.15);
    \filldraw[fill=black, line width=1.2](0,-0.5) circle (0.15);
    \filldraw[fill=gray!5, line width=1.2](0,-1.0) circle (0.15) node {$\times$};
    \filldraw[fill=gray!5, line width=1.2](0,-1.7) circle (0.15) node {$\times$};
    \end{tikzpicture}
\end{gathered}
+
\begin{gathered}
    \begin{tikzpicture}[line width=1,photon/.style={decorate, decoration={snake, amplitude=1pt, segment length=6pt}}]
    \draw[line width = 1, photon] (0,0.5) -- (1,1.5);
    \draw[line width = 1, photon] (0,-0.5) -- (1,-1.5);
    \draw[line width = 1, dashed, double] (0,-0.5) -- (0,0.5);
    \draw[line width = 1, dashed] (0,1.2) -- (0.5,1.0);
    \draw[line width = 1, dashed] (0,1.7) -- (0.8,1.3);
    \filldraw[fill=black, line width=1.2](0,0.5) circle (0.15);
    \filldraw[fill=black, line width=1.2](0,-0.5) circle (0.15);
    \filldraw[fill=gray!5, line width=1.2](0,1.2) circle (0.15) node {$\times$};
    \filldraw[fill=gray!5, line width=1.2](0,1.7) circle (0.15) node {$\times$};
    \end{tikzpicture}
\end{gathered}
+
\begin{gathered}
    \begin{tikzpicture}[line width=1,photon/.style={decorate, decoration={snake, amplitude=1pt, segment length=6pt}}]
    \draw[line width = 1, photon] (0,0.5) -- (1,1.5);
    \draw[line width = 1, photon] (0,-0.5) -- (1,-1.5);
    \draw[line width = 1, dashed, double] (0,-0.5) -- (0,0.5);
    \draw[line width = 1, dashed] (0,-1.2) -- (0.5,-1.0);
    \draw[line width = 1, dashed] (0,-1.7) -- (0.8,-1.3);
    \filldraw[fill=black, line width=1.2](0,0.5) circle (0.15);
    \filldraw[fill=black, line width=1.2](0,-0.5) circle (0.15);
    \filldraw[fill=gray!5, line width=1.2](0,-1.2) circle (0.15) node {$\times$};
    \filldraw[fill=gray!5, line width=1.2](0,-1.7) circle (0.15) node {$\times$};
    \end{tikzpicture}
\end{gathered}
+\mathcal{O}(G^3\langle Q Q\rangle ) ~.\label{eq: near zone}
\end{aligned}
\end{equation}
The one and two-loop corrections to the tidal response function have been intensively studied from explicitly loop computations \cite{Ivanov:2024sds,Caron-Huot:2025tlq}. In this paper, we focus on the calculation the first part in Eq.~\eqref{eq:farzone}  and the matching of the static and dynamical tidal Love numbers in Eq.~\eqref{eq: near zone}.

\subsection{IBP and differential equations 
\label{subsec: differential}}

Now, after we have built an integrand, we must compute the resulting integrals. This can be efficiently done using 
integration by part (IBP) identities \cite{Chetyrkin:1981qh}, which stem from the fact that total derivatives always vanish in dimensional regulariation
\begin{equation}
    \int d^D\ell \frac{\partial}{\partial \ell^\mu} \Big( \cdots\Big) = 0
\end{equation}
This allows the construction of linear relations between integrals, that enable the reduction of  complete families with arbitrary numerators to a basis of so-called \emph{master integrals}. Furthermore, the latter can be computed using differential equations arising from differentiating under the integral sign. We will now illustrate this procedure with specific one- and two-loop examples.

\subsubsection*{One-loop}
We have the following family of one-loop integrand
\begin{equation}
    g_{a_1,a_2,a_3,a_4} = \int \frac{d^D \ell}{i \pi^{D / 2}} \frac{1}{\slashed{D}_1^{a_1} D_2^{a_2} D_3^{a_3} D_4^{a_4}} ~,
\end{equation}
with propagators
\begin{equation}
    D_1=u \cdot \ell, \quad D_2=\ell^2, \quad D_3=\left(\ell+k_1\right)^2, \quad D_4=\left(\ell+k_1+k_2\right)^2 ~.
\end{equation}
Note that we use the the reversed unitarity from slashed propagator $\slashed{D}_1$ where
\begin{equation}
    \frac{1}{\slashed{D}_1^{a_1}} =
    \frac{1}{(D_1-i 0)^{a_1}} -
    \frac{1}{(D_1+i 0)^{a_1}}  \,,
\end{equation}
and the rest of the propagators have the standard Feynman prescription.

Based on Lorentz invariance, the integral can only depend on two Lorentz invariant quantites

\begin{equation}
    \omega = - u \cdot k_1 = + u \cdot k_2 ~,
\end{equation}
and the scattering angle
\begin{equation}
    x^2 = \frac{k_1 \cdot k_2}{2 \omega^2} = \sin^2 \Big( \frac{\theta}{2}\Big) ~  \Rightarrow x= \sin\(\frac{\theta}{2}\) ~.
\end{equation}
As mentioned above, IBP identities stem from the fact that
\begin{equation}
    0=\int \frac{d^D \ell}{i \pi^{D / 2}} \frac{\partial}{\partial \ell^\mu} \frac{v^\mu}{\slashed{D}_1^{a_1} D_2^{a_2} D_3^{a_3} D_4^{a_4}} ~,
\end{equation}
for any vector $v^\mu$. For instance choosing $v^\mu = u^\mu$ yields the relation
\begin{equation}
\begin{aligned}
    0 &= a_1 g_{a_1+1, a_2, a_3, a_4}-2 a_2 g_{a_1-1, a_2+1, a_3, a_4}-2 a_3 g_{a_1-1, a_2, a_3+1, a_4} \\
    & \quad -2 a_3 g_{a_1, a_2+1, a_3, a_4}-2 a_4 g_{a_1-1, a_2, a_3, a_4+1} . 
\end{aligned}
\end{equation}
After using all such relations, obtained by choosing $v^\mu \in \{u^\mu, k_1^\mu, k_2^\mu, \ell^\mu\}$, we find that there are only three master integrals
\begin{equation}
    g_1(x) = \omega^{-1+2\epsilon}g_{1010}(x,\omega) ~, \quad g_2(x) =  \omega^{1+2\epsilon}g_{1101}(x,\omega) ~, \quad g_3(x) =  \omega^{3+2\epsilon}g_{1111}(x,\omega) ~,
\end{equation}
where the dependence on $\omega$ is fixed by dimensional analysis.
Any other integral can be reduced to a linear combination of these.
The above three master integrals can be understood from their corresponding topologies where $g_1$ corresponds to the bubble topology, $g_3$ the triangle and $g_4$ the box. 
The master integrals $\vec{g} = (g_1,g_2,g_3)^T$ satisfy the Fuchsian differential equation
\begin{equation}
    \frac{d \vec{g}}{d x}=B(x, \epsilon) \vec{g}
\end{equation}
where $B$ matrix is
\begin{equation}
    B(x,\epsilon) = 
    \begin{pmatrix}
        0 & 0 & 0 \\
        0 & -\frac{2\epsilon + 1}{x} & 0 \\
        \frac{1-2\epsilon}{4x(x^2-1)} & \frac{\epsilon}{x(x^2 - 1)} & \frac{2(\epsilon + 1 -x^2)}{x(x^2 - 1)} ~.
    \end{pmatrix}
\end{equation}
There are three potential singularities in the kinematic space $x=0,\pm 1$. The $x=0$ singularity corresponds to the forward limit and $x=+1$ corresponds to the physical backward limit, where we shall impose the boundary conditions. It is worth mentioning that $x=-1$ is not the physical backward limit. The above differential equation can be made even simpler by performing the following transformation
\begin{equation}
    f_1(x) = g_{1,0,2,0}(x) ~, \quad f_2(x) = \epsilon x g_{1,1,0,1}(x) ~, \quad f_3(x) = \epsilon x^2 g_{1,1,1,1}(x) ~.
\end{equation}
which puts the differential equation into $\epsilon$-factorized form, also known as the canonical form,
\begin{equation}
\label{eq:one-loop diff}
    \frac{d\vec{f}}{dx} = \epsilon A(x) = \epsilon \[ \frac{A_0}{x} + \frac{A_{+1}}{x-1} + \frac{A_{-1}}{x+1}\] \vec{f}
\end{equation}
with $\vec{f} = (f_1,f_2,f_3)^T$ and
\begin{equation}
    A_0 = \left(\begin{array}{ccc}
0 & 0 & 0 \\
0 & -2 & 0 \\
0 & 0 & -2
\end{array}\right) ~, \quad A_{ \pm 1}=\left(\begin{array}{ccc}
0 & 0 & 0 \\
0 & 0 & 0 \\
\frac{1}{4} & \pm \frac{1}{2} & 1
\end{array}\right) ~.
\end{equation}
This form of the differential equation guarantees that the functions $f_1, f_2$ and $f_3$ have  uniform transcendental weight. In fact it shows that $f_1$ and $f_2$ are constant.

Before solving the differential equations, let us also discuss the boundary conditions we need to impose. Simply based on the fact that there is no branch cut in the physical backward limit, we require the solutions to the differential equation to be regular at $x=1$. In fact, Eq.~\eqref{eq:one-loop diff} can be solved exactly as follows
\begin{equation}
\begin{aligned}
\label{eq:one_loop_master}
    f_1(x) & = \[i \frac{e^{i\pi \epsilon}\Gamma\(\epsilon + \frac{1}{2}\)}{\sqrt{\pi}}\] \times (2\pi) ~, \\
    f_2(x) & = x^{-2\epsilon} \frac{\epsilon \pi \sec(\pi \epsilon)}{2\Gamma(1-\epsilon)} \times (2\pi) ~, \\
    f_3(x) &= \frac{1}{4} \left(1-x^2\right)^{\epsilon } \Bigg[x^{-2 \epsilon } \csc (\pi  \epsilon ) \left(-\epsilon  4 x f_2(1) \sin (\pi  \epsilon
   ) \, _2F_1\left(\frac{1}{2},\epsilon +1;\frac{3}{2};x^2\right) - \epsilon \pi  f_1(1)-\frac{2 \pi ^{3/2} f_2(1)}{\Gamma
   \left(\frac{1}{2}-\epsilon \right) \Gamma (\epsilon )}\right) \\
   & -\frac{x^2 \epsilon  f_1(1) \, _2F_1\left(\epsilon +1,\epsilon +1;\epsilon
   +2;x^2\right)}{\epsilon +1}\Bigg] ~.
\end{aligned}
\end{equation}
Furthermore, based on the exponential representation of S-matrix discussed in Sec.~\ref{subsec:exp rep} , we only need the real part of the integrals at the one-loop order. Thus, it is convenient to introduce the effective one-loop integrals $\tilde{f}_i\equiv {\rm Re} f_i,i=,1,2,3$ as
\begin{equation}
\begin{aligned}
    \tilde f_1(x) & = - \sin \pi \epsilon \frac{\Gamma(\epsilon + \frac{1}{2})}{\sqrt{\pi}} \times (2\pi) ~, \\
    \tilde f_2(x) & = x^{-2\epsilon} \frac{\epsilon \pi \sec(\pi \epsilon)}{2\Gamma(1-\epsilon)} \times (2\pi) ~, \\
    \tilde f_3(x) & = - \frac{1}{2} \pi^2 \epsilon^2 + x^{-2\epsilon} \epsilon^2 \pi^2 \Bigg( \log\Bigg(\frac{2}{x+1}\Bigg)+ \frac{1}{12}  \Bigg) + \mathcal{O}(\epsilon^3) ~.
\end{aligned}
\end{equation}

\subsubsection*{Two-loop}
For the two-loops, we have the following family of two-loop integrand
\begin{equation}
g_{a_1,a_2,a_3,a_4,a_5,a_6,a_7,a_8,a_9} = \int \frac{d^D \ell_1}{i \pi^{D/2}} \frac{d^D \ell_2}{i \pi^{D/2}} \frac{D_8^{-a_8} D_9^{-a_9}}{\slashed{D}_1^{a_1}\slashed{D}_2^{a_2} D_3^{a_3} D_4^{a_4} D_5^{a_5} D_6^{a_6} D_7^{a_7}} ~,
\end{equation}
where the seven propagators are 
\begin{equation}
    \begin{aligned}
D_1 & =u \cdot \ell_1, \quad  D_2=u \cdot \ell_2, \quad  D_3=\ell_1^2, \\
D_4 & =\ell_2^2, \quad  D_5=\left(\ell_1+k_1\right)^2, \quad D_6=\left(\ell_2+k_2\right)^2, \\
D_7 & =\left(\ell_1+\ell_2+k_1+k_2\right)^2 ~.
\end{aligned}
\end{equation}
and two irreducible numerators are
\begin{equation}
    D_8=\left(\ell_1+k_2\right)^2, \quad D_9=\left(\ell_2+k_1\right)^2 ~.
\end{equation}
After using the IBP relation, we have the following master integrals
\begin{equation}
    \begin{aligned}
g_1(x)=g_{1,1,0,0,1,1,0,0,0}(x), \quad & g_2(x)=g_{1,1,0,1,1,0,1,0,0}(x) \\
g_3(x)=g_{1,1,1,0,0,1,1,0,0}(x), \quad & g_4(x)=g_{1,1,1,1,0,0,1,0,0}(x) \\
g_5(x)=g_{1,1,1,1,0,1,1,0,0}(x), \quad & g_6(x)=g_{1,1,1,1,1,0,1,0,0}(x) \\
g_7(x)=g_{1,1,1,1,1,1,1,0,0}(x), \quad & g_8(x)=g_{1,1,2,1,1,1,1,0,0}(x) ~.
\end{aligned}
\end{equation}
Simply based on the symmetries of propagators, it is straightforward to see that 
\begin{equation}
    g_2 = g_3 ~, ~ g_5 = g_6 ~,
\end{equation}
and therefore we arrive at six master integrals in total. Similar to the one-loop case, the differential equation can be made into the canonical form via defining
\begin{equation}
    \begin{aligned}
f_1(x) & =g_{1,1,0,0,2,2,0,0,0}(x), \quad f_2(x)=f_3(x)=\epsilon g_{1,1,0,1,2,0,1,0,0}(x) \\
f_4(x)& =\epsilon(1-6 \epsilon) g_{1,1,1,1,0,0,1,0,0}(x) , \quad
f_5(x)=f_6(x)=\epsilon^2 x g_{1,1,1,1,0,1,1,0,0}(x) \\
f_7(x)& =\epsilon^2 x^2 g_{1,1,1,1,1,1,1,0,0}(x), 
\end{aligned}
\end{equation}
and
\begin{equation}
f_8(x)=\epsilon^2\left(g_{1,1,1,1,1,1,1,-1,0}(x)+g_{1,1,1,1,1,1,1,0,-1}(x)+g_{1,1,1,1,0,1,1,0,0}(x)+g_{1,1,1,1,1,0,1,0,0}(x)\right) ~.
\end{equation}
The differential equations for $\vec{f} = (f_1,f_2,f_4,f_5,f_7,f_8)^T$ can be written as
\begin{equation}
    \frac{d \vec{f}}{dx} = \epsilon \[ \frac{A_0}{x} + \frac{A_{+1}}{x-1} + \frac{A_{-1}}{x+1}\] \vec{f}
\end{equation}
with
\begin{equation}
A_0 = 
    \left(
\begin{array}{cccccccc}
 0  & 0 & 0 & 0  & 0 & 0 \\
 0 & 0 & 0 & 0  & 0 & 0 \\
 0  & 0 & -4 & 0  & 0 & 0 \\
 0  & 0 & 0 & -4 & 0 & 0 \\
 0 & 0 & 0 & 0 & -4 & 0 \\
 -2 & -4 & 1 & 0 & 32 & 2 \\
\end{array}
\right) ~, \quad 
A_{\pm 1} = \left(
\begin{array}{cccccccc}
 0  & 0 & 0 & 0  & 0 & 0 \\
 0 & 0 & 0 & 0  & 0 & 0 \\
 0 & 0 & 0 & 0& 0 & 0 \\
 0  & \pm \frac{1}{2} & \mp \frac{1}{8} & 1  & 0 & 0 \\
 0  & 0 & 0 & \pm 1 & 0 & -\frac{1}{4} \\
 0 & 2 & -\frac{1}{2} & 0 & 0 & 1 \\
\end{array}
\right) ~.
\end{equation}
The two-loop integrals cannot be solved exactly in $\epsilon$, and thus we need to adopt the perturbative expansion. Formally, given a boundary condition at a point $x=x_0$, the solution is given by a path ordered exponential, which can evaluated order by order in the $\epsilon$-expansion
\begin{equation}
    \vec f(x) = {\mathbb{P}}\text{exp}\left({\epsilon \int_{x_0}^x A(x') dx'} \right)\vec f(x_0) = \left[ 1 +\epsilon \int_{x_0}^x dx'   A(x')   + \epsilon^2  \int_{x_0}^x dx'\int_{x_0}^{x'}dx''  A(x') A(x'') + \cdots\right]\vec f(x_0)
\end{equation}
integral as
In our case we will fix the boundary condition at $x=1$, and up the order $\epsilon^2$ order, needed in this paper, the result is
\begin{equation}
\begin{aligned}
    \mathbb{P} \exp \left[\epsilon \int_1^x d A\right] = & =1+ \epsilon \Bigg\{A_0 \ln x+A_1 \log (x-1)+A_{-1}[-\log 2+\log (x+1)] \Bigg\} \\
& + \epsilon^2 \Bigg\{ A_0 A_1\left[\zeta_2- {\rm Li}_2(x)+i \pi  \log x\right]+A_1 A_0\left[-\zeta_2+ {\rm Li}_2(x)+\log x \log (1-x)\right] \\
& +A_0 A_{-1}\left[-\frac{1}{2} \zeta_2- {\rm Li}_2(-x)-\log 2 \log x\right]+A_{-1} A_0\left[\frac{1}{2} \zeta_2+ {\rm Li}_2(-x)+\log x \log (1+x)\right. \\
& +A_1 A_{-1}\left[-\frac{1}{2} \zeta_2+\frac{1}{2} \log ^2 2-\log 2 \log (1-x)+ {\rm Li}_{1,1}(x,-1)\right] \\
& +A_{-1} A_1\left[\frac{1}{2} \zeta_2-\frac{1}{2} \log ^2 2+ {\rm Li}_{1,1}(-x,-1)-i \pi \{\log 2-\log (1+x)\}\right] \\
& +\frac{1}{2} \log ^2 x A_0^2+\frac{1}{2} \log ^2(x-1) A_1^2+\frac{1}{2}[\log 2-\log (1+x)]^2 A_{-1}^2 \Bigg\} + \mathcal{O}(\epsilon^3) ~.
\end{aligned}
\end{equation}
where 
\begin{equation}
    {\rm Li}_{1,1}(x,-1) \equiv -\text{Li}_2\left(\frac{1-x}{2}\right)+\log (2) \log (1-x) +\frac{1}{2} (\zeta_2 - \log^2 2) ~.
\end{equation}
We then require that there is no branch cut at the physical backward limit at $x=1$ similar to the one-loop case. This gives us the following condition
\begin{equation}
    \begin{aligned}
    \label{eq:f1bc}
f_5(1) & =-\frac{1}{2} f_2(1)-\frac{1}{8} f_4(1) ~, \\
f_8(1) & =-2f_2(1)-\frac{1}{2} f_4(1) ~.
\end{aligned}
\end{equation}
The boundary conditions of $f_7(x)$ cannot be determined by the backward limit. Nevertheless, we can consider the forward limit at $x=0$. Note that 
\begin{equation}
    \vec{f}(x) \xrightarrow{x \rightarrow 0} x^{\epsilon A_0} \vec{f}(0) ~,
\end{equation}
where
\begin{equation}
    A_0=S_0 D_0 S_0^{-1} \quad \text { with } \quad D_0=\operatorname{diag}(-4,-4,-4,0,0,2) ~.
\end{equation}
Based on the fact that in dimensional regularization it is not possible to generate a power of $\epsilon$ with positive coefficients, we have the following additional condition
\begin{equation}
\label{eq:f0bc}
    f_7(0) = \frac{3}{16} \Big(f_1(0) + 2f_2(0) - \frac{1}{6} f_4(0) - f_8(0)\Big) ~.
\end{equation}
Our next step is to find the relation between $\vec{f}(1)$ and $\vec{f}(0)$. For ordinary differential equations, such problem is the so-called connection problem. The essential idea is to solve the differential equation from $x=0$ and $x=1$ separately and match in the overlapping region
\begin{equation}
    \vec{f}(x) = \mathbb{P} \exp \left[\epsilon \int_0^x d A\right] \vec{f}(0) = \mathbb{P} \exp \left[\epsilon \int_1^x d A\right] \vec{f}(1) ~.
\end{equation}
The connection matrix is given by
\begin{equation}
\begin{aligned}
    Z^{(+1)}(A_0,A_1,A_{-1}) & \equiv \(\mathbb{P} \exp \left[\epsilon \int_1^x d A\right] \)^{-1}  \mathbb{P} \exp \left[\epsilon \int_0^x d A\right] \\
    & = 1 + \epsilon \(- i \pi A_1 + \log (2) A_{-1}\) + \epsilon^2 \Bigg(- \frac{\pi^2}{2}A_1^2 + \frac{1}{2} \log^2 2 A_{-1}^2 - \zeta_2 [A_0,A_1]  \\
    & \quad  + \frac{1}{2} \zeta_2[A_0,A_{-1}]- \frac{1}{2} (\log^2 2 -\zeta_2)[A_1,A_{-1}] \Bigg) + \mathcal{O}(\epsilon^3) ~,
\end{aligned}
\end{equation}
where the bounary value at $x=0$ can be computed as
\begin{equation}
    \vec{f}(0) = \Big( Z^{(+1)}(A_0,A_1,A_{-1}) \Big)^{-1} \vec{f}(1) ~.
\end{equation}
Combing with the boundary condition given in Eq.~\eqref{eq:f0bc} and \eqref{eq:f1bc} yields the following condition
\begin{equation}
\begin{aligned}
\label{eq:f7_1}
    f_7(1) & = \frac{1}{16}(3f_1(1) + 12 f_2(1) - 2 f_4(1)) + \epsilon \Big(\log(2) f_2(1) - \frac{1}{4} \log(2) f_{4}(1) \Big) \\
    & \quad - \frac{1}{48} \pi^2 \epsilon^2 (f_1(1) - 24 f_3(1) + 5 f_4(1))  + \mathcal{O}(\epsilon^2) ~.
\end{aligned}
\end{equation}

Finally, according to our discussion from the exponential representation of the S-matrix, we only need the real part of the loop integrals subtracted by the iterative cuts
\begin{equation}
    \tilde f_i = {\rm Re}f_i - \frac{1}{6} f_i^{\rm cut} ~.
\end{equation}
The final results of the two-loop integral are
\begin{equation}
\begin{aligned}
    \tilde f_1 & = \frac{4 \pi ^3 \left(3 \sec ^2(\pi  \epsilon )-4\right)}{3 \Gamma \left(\frac{1}{2}-\epsilon \right)^2} ~, \\
    \tilde f_2 & = \tilde f_3 = \frac{\pi ^{5/2} 4^{\epsilon +1} \cos (2 \pi  \epsilon ) \sec (\pi  \epsilon ) \Gamma (-4 \epsilon ) \Gamma (2 \epsilon +1)}{\Gamma \left(\frac{1}{2}-3 \epsilon \right) \Gamma
   (-\epsilon )} ~, \\
   \tilde f_4 & = \frac{\pi  2^{2-4 \epsilon } x^{-4 \epsilon } \Gamma \left(\frac{1}{2}-\epsilon \right)^3 \Gamma (2 \epsilon +1)}{\Gamma \left(\frac{1}{2}-3 \epsilon \right)} ~, \\
   \tilde f_5 & = \tilde f_6 = \pi ^2 \epsilon^2 \left(-\text{Li}_2\left(x^2\right)+4 \text{Li}_2(x)+2 \log (x) \log \left(\frac{1-x}{x+1}\right)\right) + {\cal O}(\epsilon^3)~, \\
   \tilde f_7 & = -\frac{1}{3} \pi ^2 \epsilon^2 \left(\text{Li}_2\left(x^2\right)+2 \log (x) \log \left(1-x^2\right)\right) + {\cal O}(\epsilon^3) ~, \\
   \tilde f_8 & = +\frac{8}{3} \pi ^2 \epsilon \log (x) + \frac{2}{9} \pi ^2 \epsilon^2 \left(24 \text{Li}_2\left(x^2\right)-24 \log (x) \left(-2 \log \left(1-x^2\right)+\log (4 x)+\gamma_E \right)+5 \pi ^2\right) + {\cal O}(\epsilon^3) ~.
\end{aligned}
\end{equation}

\subsection{Results}

Let us now collect the results for the EFT amplitudes for scalar, electromagnetic, and gravitational waves in $D=4$, obtained by the aforementioned procedure. As discussed in Section~\ref{subsec:exp rep}, to facilitate the partial-wave transformation, instead of computing the $T$-matrix element, we shall compute the $\Delta$-matrix element.

\subsubsection*{Scalar}
 At tree level, we have
\begin{equation}
\label{eq:scalar_tree}
    i \Delta_{G} = i \frac{4 G M \pi }{x^2}
\end{equation}
At one-loop, our results are 
\begin{equation}
    i \Delta_{G^2} = i (G M \pi)^2 \omega \Big(\frac{15}{2x} - \frac{1}{2}x\Big) ~.
\end{equation}
At two-loop level, we obtain an amplitude that features a UV divergence
\begin{equation}
\begin{aligned}
    i \Delta_{G^3} & = i (G M)^3 \omega^2 \pi\Bigg[\Bigg(-\frac{15}{x} + x\Bigg) J_1(x) - \frac{16}{3 x^2} J_2(x) - 20 \log(x) + \frac{8}{\epsilon}+\frac{148}{3} -16 \log \left(\frac{4 x^2 \omega ^2}{\bar\mu^2}\right) \Bigg] \\
    & \quad + i \omega^2 (C_1^\phi + C_{\omega^20}^\phi) - 2 i C_1^\phi \omega^2 x^2 ~.
\end{aligned}
\end{equation}
where we make use of the following two functions
\begin{equation}
\label{eq:J1_J2}
    J_1(x) \equiv 2 \text{Li}_2(-x)-2\text{Li}_2(x)+ \log (x^2) \log \Big(\frac{1+x}{1-x}\Big) ~, \quad J_2(x) \equiv \text{Li}_2\left(x^2\right)+\log (x^2) \log (1-x^2)~.
\end{equation}
We also use $\bar \mu^2 = \mu^2 4\pi e^{-\gamma_E}$ in the above expression. At this order, the tidal effects enter the amplitude through Eq.~\eqref{eq:phi_tidal}. After performing the partial wave transformation formula, the results are
\begin{equation}
\begin{aligned}
\label{eq:EFT_scalar_phase_shift}
    \delta_{\ell} & = - \frac{G M \omega}{\epsilon_{\rm IR}} + (G M \omega) \(-1 - 2 \psi^{(0)}(1+\ell) + \log\(\frac{4 \omega^2}{ \bar \mu_{\rm IR}^2}\)\) \\
    & \quad + \frac{-11+ 15\ell(1 + \ell)}{(-1 + 2\ell)(1+2\ell)(3+2\ell)} \pi (G M \omega)^2 \\
    & \quad + \Bigg[ 4 \frac{-11 + 15 \ell + 15 \ell^2}{(-1+2\ell)(1+2\ell)(3+2\ell)} \psi^{(1)}(1+\ell) + \frac{4}{3}\psi^{(2)}(1+\ell) \\
    & \quad + \(\frac{4}{\ell} - \frac{4}{1+\ell} + \frac{1}{2(-1+2\ell)} - \frac{1}{2(3+2\ell)}\) \Bigg] (G m \omega)^3 ~, \quad \ell \geq 2 ~,
\end{aligned}
\end{equation}
where $\bar{\mu}_{\mathrm{IR}}^2=\mu_{\mathrm{IR}}^2 4 \pi e^{\gamma_E-1}$. From this expression, we see that there is a pole at $\ell=0$, which indicates that we are missing a local operator to cancel this divergence. We can also see this by performing partial wave transformation  at $\ell=0$
\begin{equation}
\begin{aligned}
\label{eq:EFTl0phase}
    \delta_{0} & = - \frac{G M \omega}{\epsilon_{\rm IR}} + (G M \omega) \Bigg(-1-2\gamma_E + \log\Big(\frac{4 \omega^2}{\bar \mu_{\rm IR}^2}\Big)\Bigg) + \frac{11}{3} \pi (G M \omega)^2 \\
    & \quad + \Bigg(\frac{22}{9} \pi^2 - \frac{8}{3} \zeta(3)  + \frac{2}{\epsilon} + \frac{50}{3} - 4 \log\Big(\frac{4 \omega^2}{\bar{\mu}^2}\Big) \Bigg) (G M \omega)^3 + \frac{C_{\omega^20}^\phi}{4\pi} \omega^3 ~.
\end{aligned}
\end{equation}
It is now clear that the $\ell=0$ dynamical tidal Love number is introduced to cancel the UV divergence in the EFT loop calculation. We will discuss the matching and the RG in more detail in Section~\ref{sec:matching} and Section~\ref{sec:comments on RG}. The $\ell=1$ static tidal Love also enters into the phase shift at 3PM order
\begin{equation}
\begin{aligned}
\label{eq:EFTl1phase}
    \delta_1 & = - \frac{G M \omega}{\epsilon_{\rm IR}} + (G M \omega)\(-3 + 2 \gamma_E + \log\(\frac{4 \omega^2}{\bar \mu_{\rm IR}^2}\) \) + \frac{19}{15} \pi (G M \omega)^2 \\
    & \quad + \(\frac{38}{45} \pi^2 - \frac{8}{3} \zeta(3)\) (G M \omega)^3 + \frac{C_{1}^\phi}{12 \pi} \omega^3 ~.
\end{aligned}
\end{equation}

\subsubsection*{Electromagnetic wave}

For the electromagnetic wave scatterings, we can 
decompose the exponential representation into the gauge-invariant local basis as in \eqref{eq: basis A}
\begin{equation}
    i \Delta = i \Delta_{V} V_{12} + i \Delta_H H_{12} ~,
\end{equation}
where the basis $V_{12}$ and $H_{12}$ are given in eq.~\eqref{eq: bas photon}. In $D=4$, our tree level results are 
\begin{equation}
    i \Delta_{V,G} = \frac{i G M \pi}{x^2} ~, \quad i \Delta_{H,G} = 0 ~.
\end{equation}
In general $D$, the tree level results are given by
\begin{equation}
\label{eq:photon_genD}
    i \Delta_{G} = \frac{i G M \pi}{x^2} V_{12} + \(- \frac{i(D-4)G M \pi}{(D-2) x^2}\) H_{12} ~.
\end{equation}
At one-loop order, we have
\begin{equation}
    i \Delta_{V,G^2} = \frac{1}{8} i (G M \pi)^2  \omega \frac{(3+x)(5+3x)}{x(1+x)^2} ~, \quad i \Delta_{H,G^2} = 0 ~.
\end{equation}
At two-loop order, we have
\begin{equation}
\begin{aligned}
    i \Delta_{V,G^3}& = i (G M)^3 \omega^2 \Bigg[\Bigg(-\frac{15 \pi }{4 x}+\frac{3 \pi }{2(x+1)}+\frac{ \pi }{2 (x+1)^2}+\frac{3 \pi }{2 (x-1)}-\frac{ \pi }{2(x-1)^2} \Bigg)J_1(x) \\
    & \quad + \Bigg(-\frac{4 \pi }{3 x^2}-\frac{3 \pi }{2 (x+1)}-\frac{\pi }{2 (x+1)^2}+\frac{3 \pi }{2 (x-1)}-\frac{\pi }{2 (x-1)^2} \Bigg) J_2(x) \\
    & \quad + \Bigg(-\frac{10 \pi }{3 (x+1)}-\frac{7 \pi }{6(x+1)^2}+\frac{10 \pi }{3 (x-1)}-\frac{7 \pi }{6 (x-1)^2}\Bigg) \log(x) \\
    & \quad + \frac{-7 \pi -3 \pi ^3}{6 (x+1)}-\frac{\pi ^3}{6 (x+1)^2}+\frac{7 \pi +3 \pi ^3}{6 (x-1)}-\frac{\pi ^3}{6 (x-1)^2} \Bigg] \\
    & \quad + \fft{1} {4}i\omega^2\big(C_{B,0}^{\gamma}-C_{E,0}^{\gamma}\big) ~, \\
    i \Delta_{H,G^3} & = i \frac{1}{4} \omega^2 ( C_{B,0}^\gamma + C_{E,0}^\gamma) ~,
\end{aligned}
\end{equation}
where the function $J_1(x), J_2(x)$ are given in Eq.~\eqref{eq:J1_J2}. The contribution from tidal effects given in Eq.~\eqref{eq:photon_tidal}.


We can then straightforwardly perform the partial wave transformation for the momentum space amplitudes above by using the method discussed in Section~\ref{subsec:partial_wave}. We obtain the following scattering phase shifts for $\ell \geq 2$
\begin{equation}
\begin{aligned}
\label{eq:photon_phase_shift}
    {}_1\delta_{\ell} = {}_1 \delta_{\ell,1} & = (G M \omega) \Bigg[-\frac{1}{ \epsilon_{\rm IR}} -2 \psi ^{(0)}(\ell ) -1  + \log\(\frac{4\omega^2}{\bar{\mu}_{\rm IR}^2}\) -  \frac{1+2\ell}{\ell(1+\ell)} \Bigg] \\
    & \quad + \frac{\pi  (5 \ell  (\ell +1) (3 \ell  (\ell +1)-1)-3) }{ \ell  (\ell +1) (2 \ell -1) (2 \ell +1) (2 \ell +3)} (G M \omega)^2 \\
    & \quad +  8 \Bigg[ \frac{(5 \ell  (\ell +1) (3 \ell  (\ell +1)-1)-3) \psi ^{(1)}(\ell )}{2 \ell  (\ell +1) (2 \ell -1) (2 \ell +1) (2 \ell +3)} +  \frac{1}{6} \psi^{(2)}(\ell) \\
    & \quad + \(-\frac{1}{12 \ell ^3}+\frac{13}{8 (2 \ell +1)}-\frac{5}{8 (2 \ell +3)}-\frac{1}{2 (\ell +1)}-\frac{1}{12 (\ell +1)^3} \) \Bigg] (G M \omega)^3 ~,
\end{aligned}
\end{equation}
where ${}_1\delta_{\ell}$ is associated with parity even scattering and ${}_1\delta_{\ell,1}$ corresponds to the parity odd part. The static electric and magnetic tidal Love number enters at 3PM order in the $\ell=1$ sector
\begin{equation}
\label{eq:photon_l1}
\begin{aligned}
    {}_1\delta_{1} & = (G M \omega)\( - \frac{1}{\epsilon_{\rm IR}} + 2 \gamma_E -\frac{5}{2}\) + \frac{47 \pi }{30} (G M \omega)^2 \\
    & \quad + \(-\frac{8 \zeta (3)}{3}+\frac{7}{12}+\frac{47 \pi ^2}{45} \) (G M \omega)^3 - \frac{C_{E,0}^\gamma}{6\pi} \omega^3 ~,
\end{aligned}
\end{equation}
and 
\begin{equation}
\begin{aligned}
\label{eq:photon_l11}
    {}_1 \delta_{1,1} &= (G M \omega)\( - \frac{1}{\epsilon_{\rm IR}} + 2 \gamma_E -\frac{5}{2}\) + \frac{47 \pi }{30} (G M \omega)^2 \\
    & \quad + \(-\frac{8 \zeta (3)}{3}+\frac{7}{12}+\frac{47 \pi ^2}{45} \) (G M \omega)^3 + \frac{C_{B,0}^\gamma}{6\pi} \omega^3 ~.
\end{aligned}
\end{equation}

\subsubsection*{Gravitational wave}

We adopt similar procedures in the study of gravitational wave scatterings. In $D=4$ case, we have two gauge-invariant vertices $V_{12}^2$ and $H_{12}^2$ and we can decompose the amplitude as 
\begin{equation}
    i \Delta = i \Delta_{V^2} V_{12}^2 + i \Delta_{H^2} H_{12}^2 ~.
\end{equation}
At tree level, we have the following scattering amplitude
\begin{equation}
    i \Delta_{V^2,G} = i \frac{G M \pi}{4x^2} ~, \quad i \Delta_{H^2,G} = i \frac{G M \pi}{4x^2} ~. 
\end{equation}
It is crucial to keep the recoil operator in the tree-level analysis, as it is required for gauge invariance. Dropping it results in an incorrect, non gauge-invariant amplitude.
At one-loop order, we have 
\begin{equation}
\begin{aligned}
    i \Delta_{V^2, G^2} & = i (G m \pi)^2 \omega\frac{ (x (15 x+28)+15)}{32 x (x+1)^4} ~,\\
    i \Delta_{H^2,G^2} & = 0 ~.
\end{aligned}
\end{equation}
At two-loop order, the amplitude is given by 
\begin{equation}
\begin{aligned}
    i \Delta_{V^2,G^3} & = i (G M)^3 \omega^2 \pi \Bigg[ \frac{\left(-15 x^6+7 x^4+7 x^2-15\right)}{16 x \left(x^2-1\right)^4} J_1(x) -\frac{\left(x^8+2 x^6-3 x^4+2 x^2+1\right)}{3 x^2\left(x^2-1\right)^4} J_2(x) \\
    & \quad + \frac{ \left(-134 x^6+153 x^4+72 x^2-36 \pi ^2 \left(2 x^4-3 x^2+2\right)-91\right)}{108 \left(x^2-1\right)^4} \\
    & \quad +\frac{\left(88 x^6-81 x^4+90
   x^2-81\right) \log (x)}{36 \left(x^2-1\right)^4} \Bigg] ~, \\
    i \Delta_{H^2,G^3} & = i (G M)^3 \omega^2 \pi \Bigg[\frac{(2-x^2)(10-10x^2 + x^4)}{3x^8} J_2(x) + \frac{2(60-60x^2 + 11x^4)}{9x^6} \log(x) \\
    & \quad - \frac{360 - 450 x^2 + 121 x^4}{54 x^6}\Bigg] ~,
\end{aligned}
\end{equation}
In general dimension $D$, we need to include an additional gram determinant structure $\mathcal{G}$ in \eqref{gram det}. The tree level results read 
\begin{equation}
\label{eq:grav_gen_tree}
    i\Delta_G = \( i \frac{G m \pi}{4 x^2}\) V_{12}^2 + \( i \frac{G m \pi}{4 x^2}\) H_{12}^2 + \( i \frac{2 G m \pi}{x^2}\) \mathcal{G} ~.
\end{equation}
Similar to the photon case, we get the following scattering phase shift 
\begin{equation}
\begin{aligned}
\label{eq:EFT_grav_loop}
    {}_2\delta_{\ell}& =  \left(-\frac{2 \left(2 \ell ^2+5 \ell +4\right)}{\ell  (\ell +1) (\ell +2)}-2\psi ^{(0)}(\ell -1) - 1 +  \log\(\frac{4\omega^2}{\bar{\mu}_{\rm IR}^2}\)\right) (G M \omega)\\
    & \quad + \frac{\pi  (\ell  (\ell +1) (15 \ell  (\ell +1)+13)+24)}{\ell  (\ell +1) (2 \ell -1) (2 \ell +1) (2 \ell +3)} (G M \omega)^2 \\
    & \quad + 8 \Bigg(\frac{(\ell  (\ell +1) (15 \ell  (\ell +1)+13)+24) \psi ^{(1)}(\ell -1)}{2 \ell  (\ell +1) (2 \ell -1) (2 \ell +1) (2 \ell +3)}+\frac{\psi
   ^{(2)}(\ell -1)}{6} \\
   & \quad  -\frac{7}{3 \ell ^3}-\frac{9}{\ell ^2}+\frac{407}{18 (\ell -1)}-\frac{7}{2 \ell }+\frac{371}{18 (\ell +1)}+\frac{85}{36 (\ell +2)}-\frac{225}{16 (2 \ell -1)}-\frac{1735}{36 (2
   \ell +1)} \\
   & \quad -\frac{259}{16 (2 \ell +3)}-\frac{23}{3 (\ell -1)^2}+\frac{19}{6 (\ell +1)^2}+\frac{11}{12 (\ell +2)^2}+\frac{14}{3 (\ell -1)^3}+\frac{7}{3 (\ell +1)^3}\\
   & \quad +\frac{1}{6
   (\ell +2)^3}-\frac{103}{36 (\ell -2)}+\frac{11}{12 (\ell -2)^2}-\frac{1}{6 (\ell -2)^3} \Bigg) (G M \omega)^3  ~, \\
   {}_2 \delta_{\ell,1} & = {}_2 \delta_\ell -\frac{12}{(\ell -1) \ell  (\ell +1) (\ell +2)} (G M \omega) + \frac{576}{(\ell+1)^3\ell^3 (\ell-1)^3 (\ell-2)^3} (G M \omega)^3 ~,
\end{aligned}
\end{equation}
where ${}_2\delta_{\ell}$ is associated with parity-even (polar) scattering and ${}_2\delta_{\ell,1}$ corresponds to the parity-odd (axial) part.

\section{Matching to Black Hole Perturbation Theory}
\label{sec:matching}

To extract the tidal response function of the black holes in an unambiguous and gauge invariant way, we are supposed to match the EFT scattering phase shifts directly to the ones computed by BHPT, i.e. 
\be 
\delta_\ell^{\rm EFT }= \delta_\ell^{\rm GR }\,.
\ee 
The literature with formalism to compute the BHPT phase shift is vast (see e.g., \cite{Mano:1996gn,Mano:1996mf,Mano:1996vt,Sasaki:2003xr,Ivanov:2022qqt,Saketh:2023bul,Bautista:2023sdf,Saketh:2024juq}), here we simply present the results for the scalar, photon and graviton up to $\mathcal{O}(G^3)$.

\subsection{Scalar matching}
In BHPT, the scalar phase shift result is given by
\begin{equation}
\begin{aligned}
    \delta_{\ell}^{\rm GR} & =  \[ - \psi^{(0)}(1+\ell)  - \frac{1}{2}\] (r_s \omega) + (r_s \omega) \log(2 r_s \omega) \\
    & \quad + \frac{-11+ 15\ell(1 + \ell)}{4(-1 + 2\ell)(1+2\ell)(3+2\ell)} \pi (r_s \omega)^2 \\
    & \quad + \Bigg[ \frac{1}{2} \frac{-11 + 15 \ell + 15 \ell^2}{(-1 + 2\ell)(1+2\ell)(3+ 2\ell)} \psi^{(1)}(1+\ell) + \frac{1}{6} \psi^{(2)}(1+\ell) \\
    & \quad + \( \frac{1}{2\ell} - \frac{1}{2(1+\ell)} + \frac{1}{16(-1+2\ell)} - \frac{1}{16(3+2\ell)}\) \Bigg] (r_s \omega)^3 ~, \ell \geq 1 ~,
\end{aligned}
\end{equation}
where $r_s = 2G M$ is the Schwarzschild radius. The $\ell=0$ phase shift is 
\begin{equation}
\begin{aligned}
    \delta_{0}^{\rm GR} & = \(- \frac{1}{2} + \gamma_E\) (r_s \omega) + (r_s \omega) \log(2 r_s \omega) + \frac{11}{12} \pi (r_s \omega)^2 \\
    & \quad + \( \frac{1}{2} - \gamma_E + \frac{11}{36} \pi^2 - \frac{\zeta(3)}{3} - \log(2 r_s \omega) \) (r_s \omega)^3 ~.
\end{aligned}
\end{equation}
Comparing the above expressions with the EFT results in Eq.~\eqref{eq:EFTl0phase} and \eqref{eq:EFTl1phase} yields the matching of the scalar static and dynamical Love number
\begin{equation}
\label{eq:scalar_dynamical_love}
    C_{\omega^20}^\phi(\bar{\mu})^{\overline{\mathrm{MS}}}=-4 \pi r_s^3\left[\frac{1}{4 \epsilon_{\mathrm{UV}}}+\ln \left(\bar{\mu} r_s\right)+\frac{19}{12}+\gamma_E\right] ~, \quad C^\phi_1 =0~.
\end{equation}
Similarly, the leading tidal dissipation number of black holes can be matched to the BHPT \cite{Ivanov:2024sds,Caron-Huot:2025tlq}
\begin{equation}
    C_{\omega0}^\phi = 4 \pi r_s^2 ~.
\end{equation}

\subsection{Electromagnetic matching}
The phase shift in BHPT for the scattering of photons up to $\mathcal{O}(G^3)$ are given by
\begin{equation}
\begin{aligned}
    {}_1\delta_{\ell,+}^{\rm GR} = {}_1\delta_{\ell,-}^{\rm GR} & = \( - \frac{1}{2} - \psi^{(0)}(\ell) - \frac{1+2\ell}{2\ell(1+\ell)}\) (r_s \omega) + (r_s \omega) \log(2 r_s \omega) \\
    & \quad + \frac{\pi  (5 \ell  (\ell +1) (3 \ell  (\ell +1)-1)-3) }{4 \ell  (\ell +1) (2 \ell -1) (2 \ell +1) (2 \ell +3)} (r_s \omega)^2 \\
    & \quad + \Bigg[ \frac{(5 \ell  (\ell +1) (3 \ell  (\ell +1)-1)-3) \psi ^{(1)}(\ell )}{2 \ell  (\ell +1) (2 \ell -1) (2 \ell +1) (2 \ell +3)} +  \frac{1}{6} \psi^{(2)}(\ell) \\
    & \quad \(-\frac{1}{12 \ell ^3}+\frac{13}{8 (2 \ell +1)}-\frac{5}{8 (2 \ell +3)}-\frac{1}{2 (\ell +1)}-\frac{1}{12 (\ell +1)^3} \) \Bigg] (r_s \omega)^3 ~,
\end{aligned}
\end{equation}
where the subscript $``+"$ and $``-"$ denotes the eigenvalue of parity.
Comparing the above results with the EFT loop calculation in Eqs.~\eqref{eq:photon_phase_shift},~\eqref{eq:photon_l1} and \eqref{eq:photon_l11}, i.e. ${}_1 \delta_{\ell,+1}^{\rm GR} = {}_1\delta_{\ell}^{\rm EFT}, {}_1\delta_{\ell,-1}^{\rm GR} = {}_1 \delta_{\ell,1}^{\rm EFT}$, we find that the BH spin-1 $\ell=1$ static Love number is zero, i.e.
\begin{equation}
    C_{E,0}^\gamma = C_{B,0}^\gamma = 0~.
\end{equation}

\subsection{Gravitational wave matching}
The phase shift in BHPT for the scattering of gravitons up to $\mathcal{O}(G^3)$ are given by

\begin{equation}
\begin{aligned}
    {}_2\delta_{\ell,P}^{\rm GR} & =  \left(-\frac{(2 \ell +1) \left(\ell ^2+\ell -1\right)}{(\ell -1) \ell  (\ell +1) (\ell +2)}-\psi ^{(0)}(\ell -1) - \frac{1}{2}+ \frac{3P}{(\ell-1)\ell (\ell+1) (\ell+2)}\right) (r_s \omega) + (r_s \omega) \log(2r_s \omega) \\
    & \quad + \frac{\pi  (\ell  (\ell +1) (15 \ell  (\ell +1)+13)+24)}{4 \ell  (\ell +1) (2 \ell -1) (2 \ell +1) (2 \ell +3)} (r_s \omega)^2 \\
    & \quad + (r_s \omega)^3 \Bigg(\frac{(\ell  (\ell +1) (15 \ell  (\ell +1)+13)+24) \psi ^{(1)}(\ell -1)}{2 \ell  (\ell +1) (2 \ell -1) (2 \ell +1) (2 \ell +3)}+\frac{\psi
   ^{(2)}(\ell -1)}{6} \\
   & \quad  + \frac{13}{6 \ell ^3}-\frac{9}{4 \ell ^2}+\frac{67}{4 \ell }+\frac{71}{4 (\ell +1)}+\frac{85}{36 (\ell +2)}-\frac{225}{16 (2 \ell -1)}-\frac{1735}{36 (2 \ell +1)} -\frac{259}{16
   (2 \ell +3)}\\
   & \quad +\frac{9}{4 (\ell +1)^2} \frac{11}{12 (\ell +2)^2}+\frac{13}{6 (\ell +1)^3}+\frac{1}{6 (\ell +2)^3}+\frac{85}{36 (\ell -1)}-\frac{11}{12 (\ell -1)^2}+\frac{1}{6
   (\ell -1)^3} \\
   & \quad -\frac{36 P}{(\ell -2)^3 (\ell -1)^3 \ell ^3 (\ell +1)^3}\Bigg) ~,
\end{aligned}
\end{equation}
where $P=\pm 1$ denotes the parity. This expression precisely matches the EFT  calculation in Eq.~\eqref{eq:EFT_grav_loop}, consistent with the absence of finite-size operators at this order.

\section{Comments on 
scattering and RG of high dimensional black holes}
\label{sec:comments on RG}

In this section, we comment on the scattering amplitude of scalar and electromagnetic waves on 5D black holes and gravitational wave on 7D black holes. Since we have solved the one-loop master integrals exactly in $\epsilon$ in Eq.~\eqref{eq:one_loop_master}, we are able to perform the $D$-dimensional scattering calculation through order $\mathcal{O}(G^2)$, setting the stage for a matching computation with $D>4$ BHPT. The UV logarithms of our result, allows us to extract the RG equations for the static Love numbers in $D=5,7$.

\subsection{5D scalar perturbation through $\mathcal{O}(G^2)$}
The scalar perturbation on 5D black holes have been systematically worked out in Ref.~\cite{Akhtar:2025nmt}. At tree level, the results are the same as Eq.~\eqref{eq:scalar_tree}.
At one-loop order, the results are
\begin{equation}
\begin{aligned}
    i \Delta_{G^2} & = \frac{-2 i(G M \omega)^2}{x \sqrt{1-x^2}} \operatorname{Im}\Bigg[4 \operatorname{Li}_2\left(-\frac{i x}{\sqrt{1-x^2}}\right)-4 \operatorname{Li}_2\left(\frac{i x}{\sqrt{1-x^2}}\right)-2 \operatorname{Li}_2\left(\frac{1}{2}-\frac{i x}{2 \sqrt{1-x^2}}\right) \\
    & \quad +2 \operatorname{Li}_2\left(\frac{i x}{2 \sqrt{1-x^2}}+\frac{1}{2}\right) -\log ^2\left(-1+\frac{i x}{\sqrt{1-x^2}}\right)+\log ^2\left(1+\frac{i x}{\sqrt{1-x^2}}\right) \\
    & \quad +2 i \pi \log \left(-1+\frac{i x}{\sqrt{1-x^2}}\right)+\frac{-4 \pi+4 i \log \left(2 x^2\right)}{\sin x}+3 \pi^2\Bigg] \\
& +i(G M \omega)^2\Bigg[x^2\left(\frac{8}{9} \log \left(\frac{4 \omega^2 x^2}{\bar \mu^2}\right)-\frac{32}{27}-\frac{8}{9 \epsilon_5}\right)+\frac{68}{9 \epsilon_5}-\frac{68}{9} \log \left(\frac{4 \omega^2 x^2}{\bar \mu^2}\right)-\frac{404}{27}\Bigg] \\
& + i \omega^2 (C_1^{\phi} + C_{\omega^20}^\phi) - 2 i C^\phi_1 \omega^2 x^2 ~,
\end{aligned}
\end{equation}
where we have used $D = 5 - 2\epsilon_5$. From the above expression, we immediately get the following RG coefficients
\begin{equation}
    \mu \frac{d C^\phi_{\omega^2 0}}{d \mu}=- \frac{128}{9} (G M)^2, \quad \mu \frac{d C^\phi_{1}}{d \mu}=- \frac{8}{9} (G M)^2 ~.
\end{equation}

\subsection{5D electromagnetic perturbation through $\mathcal{O}(G^2)$}
In 5D, we can also calculate the electromagnetic perturbations. At tree level, the results are given in Eq.~\eqref{eq:photon_genD}. At one-loop order, we have
\begin{equation}
\begin{aligned}
    i \Delta_{G^2} & = i (G M \omega)^2 \Bigg[ \(-\frac{1}{3 \epsilon_5} + \frac{1}{3} \log\( \frac{4\omega^2}{\bar \mu^2} \) + \Delta_{H,G^2}^{\rm finite} \) H_{12} + \( \frac{1}{3\epsilon_5} - \frac{1}{3} \log\( \frac{4 \omega^2}{\bar \mu^2}\) + \Delta_{V, G^2}^{\rm finite} \)V_{12}\Bigg] \\
    & \quad +i \frac{1}{4} \omega^2 ( C_{B,0}^\gamma + C_{E,0}^\gamma) H_{12} + i \fft{1}{4}\omega^2\big(C_{B,0}^{\gamma}-C_{E,0}^{\gamma}\big)V_{12}
\end{aligned}
\end{equation}
$\Delta_{H,G^2}^{\rm finite}(x), \Delta_{V,G^2}^{\rm finite}(x)$ will be provided in the ancillary file.
This naturally gives us the RG equation
\begin{equation}
    \mu \frac{d C_{E,0}^\gamma}{d\mu} = \frac{8}{3} (G M)^2 ~, \quad \mu \frac{d C_{B,0}^\gamma}{d\mu} = 0 ~,
\end{equation}
which agrees with the results given in \cite{Hui:2020xxx}.

\subsection{7D gravitational perturbation through $\mathcal{O}(G^2)$}

Similarly, we can perform the calculation of gravitational perturbation of 7D black holes. The tree level results are the same as Eq.~\eqref{eq:grav_gen_tree}. At one-loop order, we get
\begin{equation}
\begin{aligned}
    i\Delta_{G^2} & = i \frac{(G M \omega^2)^2}{\pi} \Bigg[ \(\frac{17}{240\epsilon_7} - \frac{17}{240}\log\(\frac{4 \omega^2}{\bar \mu^2}\) + \Delta_{G^2,H^2}^{\rm finite}(x)\) H_{12}^2 \\
    & \quad \quad +\(\frac{127}{1200\epsilon_7} - \frac{127}{1200} \log\(\frac{4 \omega^2}{\bar \mu^2}\) + \Delta_{G^2,V^2}^{\rm finite}(x)\) V_{12}^2 \\
    & \quad \quad + \(\frac{31}{45\epsilon_7} - \frac{31}{45} \log\(\frac{4 \omega^2}{\bar \mu^2}\) + \Delta_{G^2,\mathcal{G}}^{\rm finite} (x)\) \mathcal{G} \Bigg] \\
    & \quad + i\frac{1}{512} \omega^4 (3C_{T,0}^h-3C_{B,0}^h+4C_{E,0}^h)H_{12}^2 + i \frac{1}{512}\omega^4(3C_{T,0}^h+3C_{B,0}^h+4C_{E,0}^h)V_{12}^2 \\
& \quad - i \frac{1}{64}\omega^4(3C_{T,0}^h-4 C_{E,0}^h)\mathcal{G}
\end{aligned}
\end{equation}
where the explicit functions $\Delta_{H^2,G^2}^{\rm finite}(x),\Delta_{V^2,G^2}^{\rm finite}(x),\Delta_{\mathcal{G},G^2}^{\rm finite}(x)$ will be provided in the ancillary file. By comparing 
the above with Eq.~\eqref{eq:grav_ct}, we get the RG equation
\begin{equation}
    \mu \frac{dC^h_{E,0}}{d\mu} = -\frac{5024}{225\pi} (G m)^2 ~, \quad \mu \frac{d C^h_{B,0}}{d\mu} =  - \frac{448}{75\pi} (G m)^2 ~, \quad \mu \frac{d C^h_{T,0}}{d\mu} = -\frac{256}{675 \pi} (G m)^2 ~.
\end{equation}
Our above results from the on-shell loop calculation do not match the results in Ref.~\cite{Hui:2020xxx,Hadad:2024lsf} where the authors infer the RG coefficients from the static off-shell BHPT in 7D.

\section{Summary and Conclusions}

We have presented 
a systematic framework for the computation of scattering of
for massless fields off compact
gravitating objects, described 
with point-particle worldline EFT.  Such scattering is, in general, 
inelastic, which makes it 
analogous to Raman
scattering in quantum mechanics, and it can be equally used to probe the finite-size and tidal properties of the compact objects.

In this work, 
we have mostly focused 
on the elastic part of the scattering amplitude,
that is most challenging from the computational point of view. The scalar Raman computation was previously presented by us in Ref.~\cite{Ivanov:2024sds}. 
Here we generalized these results 
to the electromagnetic and 
gravitational scattering, which we computed through $O(G^3)$ and are parameterized by tidal Love-number coefficients. By projecting these amplitudes  
onto the partial wave basis
and matching with the exact solutions from black hole perturbation theory (BHPT) for the phase shift for waves in a Schwarzwald background, we explicitly match the leading static and dynamical tides.

Our photon scattering amplitudes match the 
BHPT predictions exactly, implying the vanishing of the leading electromagnetic static Love numbers. This is the first 
explicit gauge-invariant proof 
of this vanishing on shell. 

The spin-2 EFT scattering amplitude matches the black-hole  scattering phases exactly, consistent with the absence of a tidal operator at 3PM order  (the leading tidal operator will enter at 5PM). Importantly, we find that our results agree with GR only if we include the contribution 
of the recoil of the worldine, which we describe as the recoil operator of Refs.~\cite{Cheung:2023lnj,Cheung:2024byb}.  This suggests that the inclusion of the worldline recoil is required for consistency. It would be interesting to understand wether this would affect the recent computation of the dynamical gravitational response in Ref.~\cite{Combaluzier--Szteinsznaider:2025eoc}. An understanding of the recoil contributions will also be crucial in order to extend the Born series formalism of Refs.~\cite{Correia:2024jgr, Caron-Huot:2025tlq} to gravitational-wave scattering.

Finally, we also computed the spin-2 
Raman amplitudes for seven-dimensional 
black holes, which exhibit
a non-trivial RG running
already at 2PM (``one-loop order'').
Whilst the full BHPT scattering amplitude computation is not
readily available in the literature, we find a discrepancy
with the known literature results obtained from 
the matching of the off-shell 
one-point functions in EFT~\cite{Kol:2011vg,Hui:2020xxx,Hadad:2024lsf}.
The source of this discrepancy is unclear, but it
might be related the gauge-dependence
of the off-shell 
potentials, which can be contrasted with 
our gauge-invariant 
answer obtained on-shell.\footnote{This discrepancy is unlikely to be due to 
the recoil,
because it does not
affect the scheme-independent logarithmic 
terms. }
A complete BHPT  scattering amplitude computation in 7D will be likely needed 
in order to fully 
resolve the discrepancy between the off-shell and on-shell results. 

Our key result is the 
conceptualization of a 
coherent framework 
to study the tidal effects
using modern scattering 
amplitude techniques. 
In particular, 
we develop and present several 
new tools
that streamline 
the computation of gravitational 
Raman amplitudes. These include the exponential representation 
that isolates IR divergences 
in the scattering phases, 
and the partial wave generalization
to higher dimensions, allowing 
for a consistent use of dimensional regularization.

We believe that the toolkit presented here
will play a key role in future extensions of Raman scattering at higher post-Minkowskian orders. The natural targets for 
these extensions are the on-shell proof of the vanishing
of static tidal Love number
of black holes at 5PM, and the 
renormalization-group running
of the dynamical Love number
at 7PM (see~\cite{Saketh:2023bul} for matching of the logarithmic contributions using the near-far factorization). We leave these exciting 
calculations for future work.

\section*{Acknowledgements}	

We thank Simon Caron-Huot, Chih-Hao Chang, Miguel Correia, Giulia Isabella, Austin Joyce, Chia-Hsien Shen, Sasha Zhiboedov for 
useful discussions. We especially thank Yilber Fabian Bautista, Mathias Driesse, Kays Haddad and Gustav Uhre Jakobsen for comparing the results, sharing the draft and coordinating the submission of their paper \cite{Berlin}. The work of Y.Z.L is supported in part by the US National Science Foundation under Grant No. PHY- 2209997, and in part by Simons Foundation grant No. 917464.

	
\appendix

\section{Wightman functions for various states}
\label{app:wightman_func}
In this appendix, we review the properties of Wightman functions for various states.

We can consider the following positive and negative frequency wightman functions for a bosonic operator in the Heisenberg picture $O(t)$
\begin{align}
G_>(t,t') &\equiv \langle O(t)\,O(t')\rangle,\\
G_<(t,t') &\equiv \langle O(t')\,O(t)\rangle,
\end{align}
which are defined on a general state. If the state is time-translation invariant (e.g. vacuum/ground state, energy eigenstate, thermal equilibrium state and etc.), we have $G_{\gtrless}(t,t')=G_{\gtrless}(\tau)$ with $\tau=t-t'$.
The spectral function is defined by
\begin{equation}
\label{eq:spectrum_function}
\rho(\tau)\equiv \langle [O(t),O(t')]\rangle = G_{>}(\tau)-G_{<}(\tau),
\qquad
\rho(\omega)=G_{>}(\omega)-G_{<}(\omega).
\end{equation}

Let us now consider a generic density matrix for the stationary state, with $\rho =\sum_n p_n |n\rangle\langle n|$ where $H|n\rangle=E_n|n\rangle$ and $\sum_n p_n=1$.
The decomposition of Wightman functions into energy eigenstates reads
\begin{align}
G_{>}(\omega)
&=2\pi \sum_{m,n} p_n\,|O_{mn}|^2\,\delta\!\big(\omega-(E_m-E_n)\big),\\
G_{<}(\omega)
&=2\pi \sum_{m,n} p_n\,|O_{mn}|^2\,\delta\!\big(\omega+(E_m-E_n)\big),
\end{align}
where $O_{mn}\equiv \langle m|O(0)|n\rangle$.
In particular, for the ground state $|0\rangle$ one finds $G_{>}(\omega)\propto \theta(\omega)\rho(\omega)$, with only positive frequency support, and
$G_{<}(\omega)\propto \theta(-\omega)\rho(\omega)$ with only negative frequency support.

However, for the energy excited state and the thermal equilibrium, the positive (negative) frequency Wightman function will no longer be restricted to the positive (negative) definite frequencies. For example, let us consider the harmonic oscillator with $H=\omega_0(a^\dagger a+\tfrac12)$ and $x(t)=\frac{1}{\sqrt{2\omega_0}}\left(a e^{-i\omega_0 t}+a^\dagger e^{+i\omega_0 t}\right)$.
In the number operator eigenstate $|n\rangle$ we have,
\begin{align}
G_{>,n}(\tau)
&=\langle n|x(\tau)x(0)|n\rangle
=\frac{1}{2\omega_0}\Big[(n+1)e^{-i\omega_0\tau}+n\,e^{+i\omega_0\tau}\Big],\\
G_{<,n}(\tau)
&=\langle n|x(0)x(\tau)|n\rangle
=\frac{1}{2\omega_0}\Big[n\,e^{-i\omega_0\tau}+(n+1)e^{+i\omega_0\tau}\Big].
\end{align}
The Fourier transform of them in the frequency space gives
\begin{align}
G_{>,n}(\omega)
&=\frac{2\pi}{2\omega_0}\Big[(n+1)\delta(\omega-\omega_0)+n\,\delta(\omega+\omega_0)\Big],\\
G_{<,n}(\omega)
&=\frac{2\pi}{2\omega_0}\Big[n\,\delta(\omega-\omega_0)+(n+1)\delta(\omega+\omega_0)\Big].
\end{align}
We thus see that both $G_{>}$ and $G_{<}$ generally have support at $\omega=\pm \omega_0$ for $n>0$.

We now consider the thermal equilibrium state at inverse temperature $\beta$ with a Gibbs density matrix $\rho_{\beta}=e^{-\beta H}/({\rm Tr } e^{-\beta H})$,
the KMS condition implies
\begin{equation}
G_{<,\beta}(\omega)=e^{-\beta\omega}\,G_{>,\beta}(\omega).
\end{equation}
We can also write the above relation using the spectral function $\rho(\omega)$. From Eq.~\eqref{eq:spectrum_function}, we have the relation $\rho(-\omega) = \rho(\omega)$, and therefore
\begin{equation}
G_{>,\beta}(\omega)=\bigl(1+n_B(\omega)\bigr)\,\rho(\omega),
\qquad
G_{<,\beta}(\omega)=n_B(\omega)\,\rho(\omega),
\qquad
n_B(\omega)=\frac{1}{e^{\beta\omega}-1},
\end{equation}
where $n_B$ is the Bose-Einstein distribution.
For the concrete finite temperature harmonic oscillator example, we have
\begin{align}
G_{>,\beta}(\omega)
&=\frac{2\pi}{2\omega_0}\Big[(1+n_B(\omega_0))\delta(\omega-\omega_0)+n_B(\omega_0)\delta(\omega+\omega_0)\Big],\\
G_{<,\beta}(\omega)
&=\frac{2\pi}{2\omega_0}\Big[n_B(\omega_0)\delta(\omega-\omega_0)+(1+n_B(\omega_0))\delta(\omega+\omega_0)\Big].
\end{align}

In the black hole background, the Boulware vacuum $|B\rangle$ and the Hawking-Hartle vaccum $|{\rm HH}\rangle$ are the two commonly discussed states. To study classical black holes, we use the Boulware vaccum because it is the “no-particle” state annihilated by the positive frequency mode with respect to the Schwarzschild time $t$. Therefore,
\begin{equation}
    G_{>}^B(\omega) = \theta(\omega)\rho(\omega) ~, \quad G_{<}^B(\omega) = \theta(-\omega) \rho(\omega) ~.
\end{equation}
The Hartle–Hawking state describes an eternal black hole in equilibrium with a thermal bath at the Hawking temperature $\beta_H$, and hence
\begin{equation}
    G_{<,\beta_H}^{\rm HH}(\omega) = e^{-\beta_H \omega} G_{>,\beta_H}^{\rm HH}(\omega) ~.
\end{equation}

\section{Constructing partial waves}
\label{app: partial wave}

In this appendix, we provide more details of partial wave expansion in section \ref{subsec:partial_wave}. The construction closely follows \cite{Caron-Huot:2022jli}, see also \cite{Buric:2023ykg}.

\subsection{Orthogonal representations and dimensions}

We first review the orthogonal representations. A finite-dimensional irrep of SO$(d)$ is labeled by $\rho=(m_1,\cdots,m_n)$, where $n=[d/2]$ \cite{Dobrev:1977qv,Kravchuk:2017dzd}, where m’s are integers for bosonic representations and half-integers for fermionic representations, satisfying
\be
m_1\geq m_2\geq\cdots m_{n-1}\geq |m_{n}|\,.
\ee
For tensor representations, $m_i$ denotes the row lengths of the Young diagram for the irrep $\rho$. In particular, in odd $d$, $m_n$ must be positive, but it could be negative in even $d$ to indicate the chirality of the representation. We simply drop $m_i$ if it is vanishing. For example, we denote spin-$\ell$ traceless-symmetric tensor by $(\ell)$.

The dimension of a irrep $\rho$ is important to determine the normalization of partial wave expansion. We now review its formula.
\be
& {\rm dim}\,\rho=\prod_{i<j}^{d/2} \fft{p_j^2-p_i^2}{(j-1)^2-(i-1)^2}\,,\quad \text{even}\,\, d\,,\nn\\
&{\rm dim}\,\rho=\left(\prod_{i<j}^{(d-1)/2} \fft{p_j^2-p_i^2}{(j-1/2)^2-(i-1/2)^2}\right) \prod_{i}^{(d-1)/2}\fft{p_i}{i-1/2}\,,\quad \text{odd}\,\, d\,,\label{eq: dim rho}
\ee
where $p_i=m_{d/2-i+1}+i-1$ for even $d$ and $p_i=m_{(d-1)/2-i+1}+i-1/2$ for odd $d$.

For convenience, we represent tensors in irreps as index-free polynomials by using
polarization vectors $w_1,\ldots,w_n\in\mathbb{C}^d$, one for each row. The traceless and symmetry properties of a given irrep are ensured by taking these vectors to be orthogonal and defined modulo gauge redundancies \cite{Costa:2016hju}:
\begin{equation}
w_i^2 = w_i\cdot w_j = 0, \qquad
w_j \sim w_j + \#\, w_i \quad \text{for } j>i .
\label{eq:18}
\end{equation}
This ensures that the functions of $w$ must be annihilated by
$w_1\cdot \partial_{w_2}$ etc. Polynomials satisfying the gauge condition can be easily constructed by
putting vectors in the boxes of a Young tableau, where each column
represents an antisymmetrized product with $w$’s. For example, given
vectors $a^\mu,b^\mu,c^\mu\in\mathbb{C}^d$, we can define a tensor in
the $(2,1)$ representation by
\be
\myyoung{ab,c}=\left(w_1\cdot a\, w_2\cdot c-w_1\cdot c\, w_2\cdot a\right) w_1\cdot b\,.
\ee

\subsection{The normalization of partial wave expansion}

In the worldline S-matrix formalism, it is convenient to treat the worldline–particle Hilbert space as an effective one-particle Hilbert space $\mathcal{H}_p$, spanned by states $|n\rangle$ such that the momentum of the particle is $p=\omega(1,n)$. Let us keep the spin of the external particles implicit. The inner product on this Hilbert space can then be deduced from an one-particle inner product $\langle p_1|p_2\rangle=2\omega (2\pi)^{D-1}\delta^{D-1}(p_1-p_2)$, such that
\be
\fft{(2\omega)^{D-3}}{(4\pi)^{D-2}}\int_{S^{D-2}}|n\rangle\langle n|=1\,.\label{eq: completness}
\ee

In general, the Hilbert space $\mathcal{H}_p$ decomposes into a direct sum of irreducible representations $\rho$. For scalar, we only have traceless-symmetric representation $\rho=(\ell)$. Nevertheless, for photon and graviton, we have more irreps as shown in section \ref{subsec:partial_wave}. For spinning particles, we should also include polarization labels $|n,e\rangle$, where $e$ is defined by \eqref{eq: def n and e}. For each irrep $\rho$, we now choose an orthonormal basis $|i,a\rangle$ where $a$ denotes SO$(D-1)$ indices and $i$ denotes multiplicity indices \cite{Caron-Huot:2022jli}, where we have $\langle i,a|j,b\rangle=\delta^{ij}g^{ab}$ with SO$(D-1)$ invariant metric $g^{ab}$. Nevertheless, the multiplicity in our case is always $1$ in general $D$, as shown in section \ref{subsec:partial_wave}. The vertices $v^{ia}(n,e)$ are then defined by the overlap between $|i,a\rangle$ and $|n,e\rangle$
\be
\langle i,a|n,e\rangle=\left((2\omega)^{3-D}n_\rho^{(D)}\right)^{\fft{1}{2}}v^{ia}(n,e)\,,
\ee
where the normalization $n_\rho^{(D)}$ needs to be determined. We can therefore construct the projector onto $\rho$
\be
\Pi^{ij}_\rho=|i,a\rangle g_{ab}\langle j,b|\,.
\ee
The worldline S-matrix can thus be expanded in terms of these projectors
\be
S_{1{\rm w}\rightarrow 2{\rm w}}=\sum_\rho \sum_{ij}\left(S_\rho(\omega)\right)_{ji}\Pi^{ij}_\rho\,.
\ee
Taking the matrix elements of $\mathcal{M}=i(1-S)$ in states $|n,e\rangle$ yields the partial wave expansion for worldline amplitudes
\be
\langle n_2,e_2|\mathcal{S}|n_1,e_1\rangle=\sum_\rho \sum_{ij} \left(a_\rho(\omega)\right)_{ji}\langle n_2,e_2|\Pi^{ij}_\rho|n_1,e_1
\rangle=(2\omega)^{3-D}\sum_\rho n_\rho^{(D)} \sum_{ij}\left(a_\rho(\omega)\right)_{ji}\pi^{ij}_\rho\,,
\ee
where we have defined
\be
S_\rho=1+i a_\rho\,,\quad \pi^{ij}_\rho=v^{ia}(n_1,e_1)g_{ab}v^{jb}(n_2,e_2)\,.\label{eq: def partial wave}
\ee
To determine the normalization $n_\rho^{(D)}$, we note that the dimension of the irrep can be determined by the trace of the projector 
\be
\delta^{ij}{\rm dim}\,\rho={\rm Tr}\,\Pi_{\rho}^{ij}
=\fft{{\rm vol}\,S^{D-2} n_\rho^{(D)}}{(4\pi)^{D-2}}\sum_e v^{ia}(n,e)g_{ab}v^{jb}(n,e)\,,
\ee
where in the second step we insert the completeness relation \eqref{eq: completness}. Our normalization condition is
\be
\sum_e v^{ia}(n,e)g_{ab}v^{jb}(n,e)=\delta^{ij}\,.
\ee

For example, from \eqref{eq: dim rho}, we have
\be
{\rm dim}\,(\ell)=\fft{\left(2\ell+D-3\right)\Gamma\left(D+\ell-3\right)}{\Gamma\left(D-2\right)\Gamma\left(\ell+1\right)}\,.
\ee
We thus find
\be
n_{(\ell)}^{(D)}=\frac{(4\pi)^{\frac{D-2}{2}}\, (D + 2\ell - 3)\, \Gamma(D + \ell - 3)}
{\Gamma\!\left(\frac{D-2}{2}\right)\, \Gamma(\ell + 1)}\,.
\ee

\subsection{Weight-shifting operators for gluing vertices}

To glue vertices into partial waves, we need sum over the intermediate spin, i.e., $ab$ indices in \eqref{eq: def partial wave}. As we show in section \ref{subsec:partial_wave}, we perform this summation efficiently by using weight-shifting operators \cite{Karateev:2018oml,Caron-Huot:2022jli}.

A general weight-shifting operator $\mathcal{D}^{a}$ is an $\SO(D-1)$-covariant differential operator with an index $a$ for
some finite-dimensional representation of SO$(D-1)$, and it acts on a tensor in the representation $\rho$ to turn it into a new tensor in the representation with shifted weights $\rho+\de$. For our purpose, we consider weight-shifting operator $\mathcal{D}^{(h)\mu}$ that removes one box at height $h$ from a Young diagram with height $h$, which is essentially a Clebsch-Gordon coefficient for $\rho'\subset \myyoung{\ }\otimes \rho$ \cite{Karateev:2017jgd}
\be
\mathcal{D}^{(h)\mu} : \rho=(m_1,\ldots,m_h) &\to (m_1,\ldots,m_h{-}1)\equiv\rho'.
\ee
Explicitly, it is given by \eqref{todorov}, but we copy it here
\begin{equation}
\mathcal{D}^{(h)\mu_0} =
 \left(\delta^{\mu_0}_{\mu_1} - \frac{w_1^{\mu_0}}{N_1^{(h)}} \frac{\partial}{\partial w_1^{\mu_1}} \right)
 \left(\delta^{\mu_1}_{\mu_2} - \frac{w_2^{\mu_1}}{N_2^{(h)}}\frac{\partial}{\partial w_2^{\mu_2}}  \right)\cdots
\left(\delta^{\mu_{h-1}}_{\mu_{h}} - \frac{w_h^{\mu_{h-1}}}{N_h^{(h)}{-}1}\frac{\partial}{\partial w_h^{\mu_{h}}}  \right)
\frac{\partial}{\partial w_{h\mu_{h}}}\,,
\end{equation}
where $N_i^{(h)}=D-2+m_i+m_h-i-h$. For $h=1$, as expected it gives Todorov/Thomas operator that
acts on traceless symmetric tensors \cite{Dobrev:1975ru}
\be
\mathcal{D}^{(1)\mu}= \fft{\partial}{\partial w_\mu}- \fft{w^\mu}{D-4+2\ell} \fft{\partial}{\partial w_1}\cdot \fft{\partial}{\partial w_1}\,.
\ee
We have already outlined the procedure to use weight-shifting operators to construct partial waves. Here, we comment on consistent properties that they have to obey \cite{Caron-Huot:2022jli}.

First, $\mathcal{D}^{(h)\mu}$ must preserve the gauge constraints: for all $i<j$, $w_i{\cdot}\partial_{w_j}\mathcal{D}^{(h)\mu}X=0$ if $X$ satisfies the same constraint. Second, $\mathcal{D}^{(h)\mu}$ has to send traces to traces, where the `traces'' denotes index contractions in strictly gauge-invariant polynomials, for example it sums over an index $\mu$ denoting a unit vector in a particular direction
\be
\sum_{\mu=1}^{d} \myyoung{ac,b\mu,\mu}.
\ee
These properties are crucial as they determine $\cD^{(h)\mu}$ up to an overall normalization, which can be fixed by considering traces on height-$h$ columns. This is precisely how we determine $N_i^{(h)}$.

\subsection{Verifying the orthonormality of partial waves}

We now verify the orthonormality of the partial waves we construct. The explicit expressions for the partial waves are provided in the ancillary file; here we simply quote the integral identities, which have been explicitly checked.
\be
& \int_{-1}^1 dx (1-x^2)^{\fft{D-4}{2}}\sum_{e_i,e_i^\ast }\pi_{s,\ell_1}^{(\ell)}(e_i)\pi_{s,\ell_2}^{(\ell)}(e_i^\ast)=\frac{2^{D-3} \Gamma \left(\frac{D}{2}-1\right)^2 \Gamma (\ell+1)}{(D+2 \ell-3) \Gamma (D+\ell-3)}\delta_{\ell \ell'}\,,\nn\\
&\int_{-1}^1 dx (1-x^2)^{\fft{D-4}{2}}\sum_{e_i,e_i^\ast }\pi_{s,\ell_1}^{(\ell,1)}(e_i)\pi_{s,\ell_2}^{(\ell,1)}(e_i^\ast)=\frac{2^{D-3} (\ell+1) \Gamma \left(\frac{D}{2}-1\right)^2 \Gamma (\ell)}{(D-3) (D+\ell-3) (D+2 \ell-3) \Gamma (D+\ell-4)} \delta_{\ell \ell'}\,,\nn\\
& \int_{-1}^1 dx (1-x^2)^{\fft{D-4}{2}}\sum_{e_i,e_i^\ast }\pi_{s,\ell_1}^{(\ell,2)}(e_i)\pi_{s,\ell_2}^{(\ell,2)}(e_i^\ast)=\frac{\sqrt{\pi } \Gamma \left(\frac{D}{2}-2\right) \Gamma (D-2) \Gamma (\ell+2)}{2 (\ell-1) (D+\ell-2) (D+2 \ell-3) \Gamma \left(\frac{D}{2}+\frac{1}{2}\right) \Gamma (D+\ell-4)} \delta_{\ell \ell'}\,,\label{eq: normorth partial wave}
\ee
where $s$ denotes the scalar, photon, and graviton. For the $(\ell)$ irrep all three are present, while in the $(\ell,1)$ irrep only the photon and graviton appear, and in the $(\ell,2)$ irrep only the graviton appears.

\section{RG coefficients in 4D}
\label{app:RG_coefs}
In this appendix, we discuss the RG running behavior of tidal operators, i.e. the anomalous dimensions. Fundamentally speaking, the RG running is coming from the logarithmic corrections due to gravitational nonlinear interactions in the bulk. The most general form of the tidal RG running has been proposed in Ref.~\cite{Caron-Huot:2025tlq} which has the constant, linear and quadratic piece
\begin{equation}
\begin{aligned}
    \mu \frac{dG_{R,\ell}(\omega,\mu)}{d\mu} & = (GM)^{2\ell+1}\(\sum_{n=1}\gamma_{\ell,n}^{(0)}(GM \omega)^n\) + \(\sum_{n=1} \gamma_{\ell,n}^{(1)} (G M \omega)^n\) G_{R,\ell}(\omega,\mu) \\
    & \quad + \omega^{2\ell+1} \(\sum_{n=1} \gamma_{\ell,n}^{(2)}(G M \omega)^n\) G_{R,\ell}(\omega,\mu)^2 ~.
\end{aligned}
\end{equation}
The above RG equation explicitly depends on the perturbation frequency which is different from the standard form appearing in the QFT textbook \cite{Peskin:1995ev,Schwartz:2014sze}
\begin{equation}
    \mu\frac{d O_{i}}{d\mu} = \gamma_{ij}(g(\mu)) O_{ij} ~,
\end{equation}
where $O_i$ represents the general operator and $g(\mu)$ is the scale dependent coupling in the theory. This is because from the worldline point of view, the running was induced by the nonlinear interactions in the gravitational environment. Therefore, this is not the anomalous dimension in the closed system, but rather the anomalous dimension in the open system. 
We show explicitly some of the RG coefficients that have been previously computed in \cite{Caron-Huot:2025tlq,Ivanov:2025ozg}. For scalar perturbations, most of the known results at $\ell=0,1,2$ are given in Ref.~\cite{Caron-Huot:2025tlq}. We list the results for $\ell=0$ sector here
\begin{equation}
\begin{aligned}
    \gamma_{0,1}^{(0)} &=0 ~, \quad \gamma_{0,2}^{(0)} = -32\pi ~, \quad \gamma_{0,3}^{(0)} = 0 ~, \quad \gamma_{0,4}^{(0)} = -288\pi ~, \quad \gamma_{0,5}^{(0)} = 0 ~, \quad \gamma_{0,6}^{(0)} = -\frac{2081632 \pi }{675} ~, \\
    \gamma_{0,1}^{(1)} & = 0 ~, \quad \gamma_{0,2}^{(1)} = -\frac{44}{3} ~, \quad \gamma_{0,3}^{(1)} =0 ~, \quad \gamma_{0,4}^{(1)} = -\frac{14236}{135} ~, \quad \gamma_{0,5}^{(1)} = 0 ~, \quad \gamma_{0,6}^{(1)} = -\frac{44658304}{42525} ~, \\
    \gamma_{0,1}^{(2)} & = -\frac{1}{\pi } ~, \quad \gamma_{0,2}^{(2)} = 0 ~, \quad \gamma_{0,3}^{(2)} = -\frac{28}{3 \pi } ~, \quad \gamma_{0,4}^{(2)} = 0 ~, \quad \gamma_{0,5}^{(2)} = -\frac{62701}{810 \pi } ~, \quad \gamma_{0,6}^{(2)} = 0 ~, \\
    \gamma_{0,7}^{(2)} & =-\frac{3353908729}{3827250 \pi } ~.
\end{aligned}
\end{equation}
In addition to these results, we have proved in Ref.~\cite{Ivanov:2025ozg} that the universal part of the linear in $F$ anomalous dimension coincides with the ``renormalized" angular momentum in black hole perturbation theory, i.e. 
\begin{equation}
    \gamma_{\ell,n}^{(1)} = 2(\nu_{\ell,n}-\ell)~, \quad n \leq2\ell+2 ~,
\end{equation}
which gives a prediction for many previously unknown RG coefficients.
For scalar $\ell=3$ perturbations, this corresponds to
\begin{equation}
\begin{aligned}
    \gamma_{3,1}^{(1)} & = 0, ~, \quad \gamma_{3,2}^{(1)} = -\frac{676}{315} ~, \quad \gamma_{3,3}^{(1)} = 0 ~, \quad \gamma_{3,4}^{(1)} = -\frac{297521684}{343814625} ~, \quad \gamma_{3,5}^{(1)} =0 ~, \quad \\
    \gamma_{3,6}^{(1)} & = -\frac{4034903368173956}{4878445881684375} ~, \quad \gamma_{3,7}^{(1)} = 0 ~, \quad \gamma_{3,8}^{(1)} = -\frac{1157205412079291092802}{1064940343742290640625} ~.
\end{aligned}
\end{equation}
For gravitational $\ell=2,3$ perturbations, this leads to the following RG coefficients
\begin{equation}
\begin{aligned}
    {}_2\gamma_{2,1}^{(1)} & = 0~, \quad {}_2\gamma_{2,2}^{(1)} = -\frac{428}{105} ~, \quad {}_3\gamma_{2,3}^{(1)} = 0 ~, \quad {}_2 \gamma_{2,4}^{(1)} = -\frac{6780932}{1157625} ~, \quad {}_2 \gamma_{2,5}^{(1)} = 0 ~, \quad {}_2 \gamma_{2,6}^{(1)} = -\frac{3024789542476}{140390971875} ~, \\
    {}_2 \gamma_{3,1}^{(1)} & = 0 ~, \quad {}_2 \gamma_{3,2}^{(1)} = -\frac{52}{21} ~, \quad {}_2 \gamma_{3,3}^{(1)} = 0 ~, \quad {}_2\gamma_{3,4}^{(1)} = -\frac{43684}{33957} ~, \quad  {}_2 \gamma_{3,5}^{(1)} = 0 ~, \quad {}_2 \gamma_{3,6}^{(1)} = -\frac{762830658152}{481821815475} ~, \\
    {}_2 \gamma_{3,7}^{(1)} &= 0 ~, \quad {}_2 \gamma_{3,8}^{(2)} = -\frac{94654540068335732}{35059764403038375} ~,
\end{aligned}
\end{equation}
that may be useful to perform the resummation in the GW waveform.

\section{Details of the ancillary file}
\label{app: anc}

We anticipate that the partial waves constructed in this paper may serve future research on gravitational Raman scattering and perhaps other topics. Therefore, we have prepared the \texttt{Partialwave$\_$Worldline.m} file, which contains the complete information used to set up the worldline partial waves for the scalar, photon, and graviton. The file uses the notation $d$ for the spacetime dimension (denoted as $D$ in the main text) and $j$ for the spin ($\ell$ in the main text). The file contains:
\begin{itemize}
    \item The local basis \texttt{localbasis[A]} and \texttt{localbasis[h]}, used in this paper to encode the polarization structures \eqref{eq: basis A} and \eqref{eq: basis h}.
    \item The lists \texttt{vert[A]} and \texttt{vert[h]} of three-point couplings between the photon or graviton, the worldline, and a massive state, written in the Young Tableau notation of Section 3.3 and divided by the normalization factor. Here, \texttt{sc[a,b]} denotes $a\cdot b$, and \texttt{p[i], e[i],w[i]} denote $p_i, e_i, w_i$.
    \item The partial waves \texttt{partialwaves[S]}, \texttt{partialwaves[A]}, and \texttt{partialwaves[h]}, providing lists of partial waves with a given irreducible representation and organized in terms of the local basis with the ordering in \texttt{localbasis[]}. \texttt{pj[x,n]} denotes $\partial_x^n P_J(x)$.
    \item The lists \texttt{norms[S]}, \texttt{norms[A]}, and \texttt{norms[h]} of orthonormal normalizations for given irreducible representations \eqref{eq: normorth partial wave}.
    \item The function \texttt{$\rho$dim[irrep]}, which provides the dimension of the irreducible representations used in this paper.
    \item The formula \texttt{nrep[irrep]} for the normalization of the partial wave expansion $n_\rho^{(D)}$, and \texttt{Vol[d]}, which denotes the $d$-dimensional volume.
\end{itemize}

In the file \texttt{highD\_const.m}, we provide the explicit functions $\Delta_{G^2,V}^{\rm finite}, \Delta_{G^2,H}^{\rm finite}, \Delta_{G^2,V^2}^{\rm finite}, \Delta_{G^2,H^2}^{\rm finite}$ and $\Delta_{G^2,\mathcal{G}}^{\rm finite}$ appearing in the finite part of the scattering upon high dimensional black holes.

\bibliographystyle{JHEP}
\bibliography{long}

\end{document}